\documentclass[aps,prl,reprint,twocolumn,superscriptaddress,longbibliography,nofootinbib,floatfix,showpacs]{revtex4-1}
\usepackage{epsfig}
\usepackage{amsmath,amssymb,amsfonts}
\usepackage{mathrsfs}
\usepackage{bbm}
\usepackage{slashed}
\usepackage{graphicx}
\usepackage{verbatim}
\usepackage[T1]{fontenc}
\usepackage[utf8]{inputenc}
\usepackage[colorlinks]{hyperref}
\usepackage{ifthen}
\usepackage{forloop}
\usepackage[english]{babel}
\usepackage{mathbbol}
\usepackage[usenames]{color}
\usepackage[normalem]{ulem}

\definecolor{darkgreen}{rgb}{0.2,0.6,0}
\definecolor{lightblue}{rgb}{0,0.5,0.8}
\definecolor{lightred}{rgb}{0.8,0.2,0.2}
\definecolor{darkorange}{rgb}{1,0.549,0}
\definecolor{brown}{rgb}{0.609, 0.164, 0.164}

\newcommand{\cA}{{\mathcal A}}

\newcommand{\GR}{{\small GR}}

\newcommand{\eg}{{\textit{e.g.}}}

\graphicspath{{figures/}}

\newcommand{\be}{\begin{equation}}
\newcommand{\ee}{\end{equation}}
\newcommand{\ba}{\begin{eqnarray}}
\newcommand{\ea}{\end{eqnarray}}

\newcommand{\mans}{\ensuremath{\mathfrak{s}}}
\newcommand{\mant}{\ensuremath{\mathfrak{t}}}
\newcommand{\manu}{\ensuremath{\mathfrak{u}}}

\newcommand{\ressym}{\hat{\Gamma}}

\newcommand{\modelname}{model}

\allowdisplaybreaks

\begin{document}

\title{Finite Quantum Gravity Amplitudes -- no strings attached}
\author{Tom Draper\,\href{https://orcid.org/0000-0002-6895-894X}{\protect \includegraphics[scale=.07]{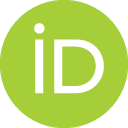}}}
\email[]{t.draper@student.ru.nl}
\affiliation{
Institute for Mathematics, Astrophysics and Particle Physics (IMAPP),\\
Radboud University Nijmegen, Heyendaalseweg 135, 6525 AJ Nijmegen, The Netherlands
}
\author{Benjamin Knorr\,\href{https://orcid.org/0000-0001-6700-6501}{\protect \includegraphics[scale=.07]{ORCIDiD_icon128x128.png}}}
\email[]{bknorr@perimeterinstitute.ca}
\affiliation{
Perimeter Institute for Theoretical Physics, 31 Caroline St. N., Waterloo, ON N2L 2Y5, Canada
}
\author{Chris Ripken\,\href{https://orcid.org/0000-0003-2545-5047}{\protect \includegraphics[scale=.07]{ORCIDiD_icon128x128.png}}}
\email[]{aripken@uni-mainz.de}
\affiliation{
Institute of Physics (THEP), University of Mainz, Staudingerweg 7, 55099 Mainz, Germany
}
\author{Frank Saueressig\,\href{https://orcid.org/0000-0002-2492-8271}{\protect \includegraphics[scale=.07]{ORCIDiD_icon128x128.png}}}
\email[]{f.saueressig@science.ru.nl}
\affiliation{
Institute for Mathematics, Astrophysics and Particle Physics (IMAPP),\\
Radboud University Nijmegen, Heyendaalseweg 135, 6525 AJ Nijmegen, The Netherlands
}

\begin{abstract}
We study the gravity-mediated scattering of scalar fields based on a parameterisation of the Lorentzian quantum effective action. We demonstrate that the interplay of infinite towers of spin zero and spin two poles at imaginary squared momentum leads to scattering amplitudes that are compatible with unitarity bounds, causal, and scale-free at trans-Planckian energy. Our construction avoids introducing non-localities or the massive higher-spin particles that are characteristic in string theory. 
\end{abstract}
%

%
\maketitle

\textit{Introduction.}---Relativistic quantum field theories, including the Standard Model of particle physics, have been extremely successful in predicting the outcome of particle physics experiments.
Gravitational physics, including the motion of planets and gravitational waves, are very well-described via the laws of classical general relativity (\GR{}).
However, promoting \GR{} to a quantum field theory results in a theory that is perturbatively non-renormalisable \cite{'tHooft:1974bx, Goroff:1985sz, Goroff:1985th}.
Physically, this deficit manifests itself in amplitudes describing the gravity-mediated scattering of particles.
At tree level these amplitudes diverge quadratically in the centre-of-mass energy.
Adding loop-corrections aggravates these divergences \cite{Anber:2011ut}.
Nevertheless, \GR{} is a completely predictive and unitary effective field theory up to the Planck scale $M_{\rm Pl} \simeq 10^{19}$~GeV \cite{Donoghue:1993eb, Donoghue:1994dn,Aydemir:2012nz}.

Requiring that the effective field theory can be completed into a fundamental theory valid on all scales generically puts constraints on the scattering amplitudes by requiring unitarity and causality \cite{Adams:2006sv, Camanho:2014apa, Bellazzini:2015cra, Chandrasekaran:2018qmx}.
These constraints constitute challenges for many quantum gravity programs.
For instance, Stelle gravity \cite{Stelle:1976gc, Stelle:1977ry} resolves the growth of the amplitudes at high energy at the expense of unitarity or causality \cite{Anselmi:2019nie, Donoghue:2019ecz}, while Lee-Wick type gravity models \cite{Tomboulis:1977jk, Tomboulis:1980bs,Anselmi:2018kgz, Anselmi:2018tmf, Anselmi:2018bra, Anselmi:2020tqo} break causality at sub-Planckian scales (also see \cite{Shapiro:2015uxa,Salles:2018ccb} for a related discussion of massive ghost modes in non-local gravity models).
One way of resolving the divergent amplitudes found in \GR{} introduces an infinite tower of massive higher-spin resonances corresponding to new particles.
This leads to the well-known Veneziano and Virasoro-Shapiro amplitudes \cite{Veneziano:1968yb, Virasoro:1969me, Shapiro:1970gy,Alonso:2019ptb}, see \cite{Green:1987sp} for a review.
However, at this point one leaves the framework of quantum field theory, introducing a vast range of 
new physics including the string theory landscape comprising an exponentially large number of vacuum states \cite{Douglas:2003um, Taylor:2015xtz}.
This raises the natural question whether it is possible to accommodate all consistency requirements on the scattering amplitudes also within the framework of a  quantum field theory.
In this Letter we demonstrate that the answer to this question is affirmative: the interplay of an infinite tower of \emph{massless} (Lee-Wick type) poles which asymptotically approach a Regge-type behaviour leads to gravity-mediated scattering amplitudes which are free from high-energy divergences and meet the unitarity and causality constraints.
Our explicit example serves as a proof of principle, highlighting the essential features of such a construction, while being easily generalisable.
We stress that our work focuses on the high-energy behaviour of the scattering amplitudes only.
In particular, infrared divergences related to the massless nature of the graviton will not be discussed.

\textit{The quantum effective action.}---Our exposition focuses on the gravity-mediated scattering of two distinguishable, massless scalar particles.
The starting point is the quantum effective action $\Gamma$ in a Lorentzian signature spacetime.
By definition this action includes \emph{all quantum corrections} so that scattering amplitudes can be constructed from tree-level Feynman diagrams involving the vertices and propagators obtained from $\Gamma$.
Since gravitational interactions are long-ranged, it is expected that $\Gamma$ contains local as well as non-local terms.

In this work we will not attempt to connect $\Gamma$ to a specific microscopic quantum gravity model.
Instead we parameterise the quantum effective action as
\begin{equation}\label{eq:Gamma}
\begin{aligned}
 \Gamma &\simeq \frac{1}{16\pi G_N} \int \sqrt{-g} \bigg[ \frac{1}{2} \phi \Delta \phi + \frac{1}{2} \chi \Delta \chi	+ \frac{1}{4} f_{\phi\chi} \, \phi^2 \chi^2 \\
 & \qquad - R - \frac{1}{6} R \, f_R(\Delta) \, R + \frac{1}{2} C_{\mu\nu\rho\sigma} \, f_{C}(\Delta) \, C^{\mu\nu\rho\sigma} \bigg] \, .
\end{aligned}
\end{equation}
Here  $\Delta \equiv - g^{\mu\nu} D_\mu D_\nu$ is the d'Alembert operator constructed from the Lorentzian spacetime metric $g_{\mu\nu}$, and $R$ and $C_{\mu\nu\rho\sigma}$ denote the Ricci scalar and Weyl tensor constructed from $g_{\mu\nu}$, respectively.
The two form factors $f_{R}$ and $f_{C}$ capture corrections to the graviton propagator obtained from the Einstein-Hilbert action and their numerical pre-factors have been chosen for later convenience.
The scalar fields $\phi,\chi$ are taken to be massless which is well-justified when considering particle scattering at trans-Planckian energy.
Their self-interaction includes the form factor $f_{\phi\chi}(\{-D_{i\mu} D_j^\mu \}_{i<j})$ which depends on the contracted covariant derivatives $D_i^\mu$ acting on the $i$-th scalar field in the expression.
The function $f_{\phi\chi}$ is symmetric under exchanging $D_1^\mu \leftrightarrow D_2^\mu$ and $D_3^\mu \leftrightarrow D_4^\mu$.
Notably, the action \eqref{eq:Gamma} encodes \emph{the most general form of the gravitational propagator} compatible with invariance under general coordinate transformations when considering gravitons propagating in a flat Minkowski space \cite{Barvinsky:1990up}.
For vanishing form factors the action \eqref{eq:Gamma} coincides with the Einstein-Hilbert action supplemented by two minimally coupled scalar fields.
In general, quantum corrections will lead to non-vanishing form factors.
The prototypical example is the treatment of gravity as an effective field theory where $f_{R,C} \simeq \log(\Delta/\mu^2)$ \cite{'tHooft:1974bx, Donoghue:1993eb, Donoghue:1994dn}.

\textit{Specifying the form factors.}---
In general, we require that the inclusion of form factors a) does not introduce new poles in the propagator at real squared momentum and b) yields amplitudes that are bounded and scale-free for large squared momentum.
These requirements can be realised as follows.\footnote{A more complete discussion will be given elsewhere \cite{Draper:2020knh}.} 
In the gravitational sector, we take
\begin{equation}\label{infiniteres}
				f_{R,C}(\Delta)
		=
				c_{R,C} \, G_N	\, \tanh \left(	c_{R,C} \,	G_N	\, \Delta	\right)
		\, .
\end{equation}
Here $G_N \Delta$ is a dimensionless combination which measures the momentum of the graviton in units of the Planck mass $M_{\rm Pl}^2 \equiv G_N^{-1}$.
The two numerical parameters $c_{R}, c_{C} > 0$ control the position of the imaginary poles in the propagator.
The construction is completed by the matter form factor $f_{\phi\chi}$ whose contribution to the scattering amplitudes is detailed in \eqref{matterff1} and \eqref{matterpara} below.
Notably, the action \eqref{eq:Gamma} is local in the sense that it involves only finitely many derivatives as the (generalised) momentum-scale is sent to infinity.

The flat space graviton propagator resulting from the action \eqref{eq:Gamma} is obtained in the standard way.
Denoting the graviton momentum by $p^2$, the gauge-fixed graviton propagator reads
\begin{equation}\label{IRMprop}
 G(p^2) \propto \Pi_\text{TT} \frac{1}{p^2(1+p^2 \, f_C(p^2))} -2 \Pi_0 \frac{1}{p^2(1+p^2 \, f_R(p^2))} \, ,
\end{equation}
where the projectors $\Pi_{\text{TT},0}$ project onto the spin two and zero modes.
For real squared momentum, the propagator possesses simple poles at $p^2 = 0$ corresponding to the pole structure familiar from \GR{}.
Otherwise it is regular on the entire real axis and falls off like $p^{-4}$ for asymptotically large momenta.

\begin{figure}
	\includegraphics[width=0.5\columnwidth]{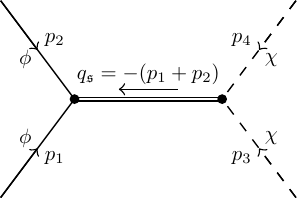}
	\caption{Feynman diagram encoding the amplitude $\mathcal{A}_\mans$ associated with $\mans$-channel scattering of two distinct scalar particles mediated by graviton exchange. All vertices and propagators are effective quantities that include all quantum corrections.}
	\label{fig:feynman}
\end{figure}

\textit{Finite scattering amplitudes.}--- Scattering amplitudes are conveniently parameterised in terms of the Mandelstam variables $\mans \equiv (p_1 + p_2)^2$, $\mant \equiv (p_1 + p_3)^2$, and $\manu \equiv (p_1 + p_4)^2$, subject to the (massless) relation $\mans + \mant + \manu = 0$, cf.\ \autoref{fig:feynman} for our convention of the particle momenta.
Here $\mans$ encodes the energy of the scattering process in the centre-of-mass frame.

In order to exhibit the basic mechanism underlying our construction we first consider the graviton-mediated scattering process $\phi\phi \to \chi\chi$, setting $f_{\phi\chi} = 0$.
As the incoming and outgoing particles are distinguishable, the amplitude is determined by the single $\mans$-channel tree-level Feynman diagram \autoref{fig:feynman}.
Imposing on-shell conditions for the external legs, the resulting amplitude $\cA_\mans$ is
\begin{equation}\label{scats}
\begin{aligned}
\mathcal{A}_\mans & = 
\frac{4\pi G_N}{3}	\left[
\frac{\mans}{(1+ \mans f_R(\mans))} - \frac{\mant^2-4 \mant \manu + \manu^2}{\mans (1+\mans f_C(\mans))}
\right]
\\ 
& =
\frac{4\pi G_N}{3} \mans^2	\bigg[
 G_R(\mans) -	P_2(\cos\theta) G_C(\mans)
\bigg] \, .
\end{aligned}
\end{equation}
Here $\theta$ is the scattering angle in the centre-of-mass frame, $P_2(x) \equiv (3x^2-1)/2$ is the second Legendre polynomial, and $G_{R,C}(p^2) \equiv [p^2 (1+p^2 f_{R,C}(p^2))]^{-1}$.
We performed the  calculation with arbitrary gauge parameters to explicitly show that \eqref{scats} is gauge-independent.

Starting from the amplitude \eqref{scats}, it is straightforward to compute the partial-wave amplitudes $a_j(\mans)$ of spin $j$,
\be\label{pwa}
a_j(\mans) \equiv \frac{1}{32\pi} \int_{-1}^1 \text{d}\cos\theta \, P_j(\cos\theta) \, \cA_\mans(\mans,\cos\theta) \, .
\ee
Evaluating the integral gives
\be\label{eq:partialwaves}
\begin{split}
	a_0(\mans) =  \, \frac{G_N}{12} \, \mans^2 \, G_R(\mans), \; \; \;  
	a_2(\mans)  = & \, - \frac{G_N}{60} \, \mans^2 \, G_C(\mans) \, .
\end{split}
\ee
All other partial-wave amplitudes vanish.
This coincides with the idea that we have just modified the propagation of the gravitational degrees of freedom without introducing propagating degrees of freedom of higher spin.
For $f_{R,C} = 0$, eq.\ \eqref{eq:partialwaves}  reduces to the amplitudes obtained from \GR{},
\be\label{Asgr}
	a_0^{\rm GR}(\mans) =  \, \frac{G_N}{12} \, \mans \, ,  \quad 
a_2^{\rm GR}(\mans)  =  \, - \frac{G_N}{60} \, \mans \, ,
\ee
showing that in this case the partial-wave amplitudes grow linearly with the energy transfer $\mans$.

\textit{Constraints on the amplitude.}---In order to be viable,  a scattering amplitude has to obey bounds originating from different physics requirements:
\begin{enumerate}
	\item[1)] \emph{Unitarity:}
		The partial-wave amplitudes describe the overlap of in- and out-states in the scattering process.
		Imposing that the probability of going from an in-state to a specific out-state does not exceed $1$ bounds the absolute value of the partial-wave amplitudes,
	\be\label{eq:ajbound}
 \left| a_j(\mans) \right| \le 1 \, , \quad \forall j \ge 0 . 
	\ee
		These constraints imply that the construction obeys the Froissart bound\footnote{The Froissart bound is a consequence of the optical theorem relating the imaginary part of the forward scattering amplitude to the total cross section \cite{Froissart:2010}. Notably, the optical theorem is a genuine quantum relation, requiring the presence of loop-corrections in the amplitude part, see \cite{Anselmi:2019xac} for a pedagogical discussion. This indicates that the form factors in \eqref{eq:Gamma} necessarily include non-analytic terms in order to fulfil the optical theorem. Determining these structures is beyond the scope of the present work and we will work with the simpler condition \eqref{eq:ajbound}. We expect that they will not lead to difficulties in the construction.}  \cite{Froissart:1961ux, Froissart:2010}, stating that the total cross section $\sigma_{\rm tot}$ cannot grow faster than $\log^2 \mans$. 

	\item[2)] \emph{Causality:}
		For large $\mans$ at fixed $\mant$ (corresponding to forward scattering), causality implies that the amplitude $\cA(\mans,\mant)$ must be polynomially bounded growing slower than $\mans^2$ \cite{Camanho:2014apa}.

	\item[3)] \emph{Cerulus-Martin bound} \cite{Cerulus:1964cjb,Epstein:2019zdn}:
	For large $\mans$ and fixed scattering angle, causality implies that the amplitude cannot fall faster than $e^{-\sqrt{\mans} \ln \mans}$.
\end{enumerate}
Clearly, the partial-wave amplitudes found in \GR{}, eq.\ \eqref{Asgr}, violate condition 1) for $\mans \gtrsim M_{\rm Pl}^2$.
This has triggered many speculations about the physics encoded in the amplitudes at trans-Planckian scales including, \eg{}, the formation of black holes as intermediate states in the scattering process \cite{Giddings:2009gj} or a classical self-completion of \GR{} \cite{Dvali:2010jz}.

\begin{figure}
	\includegraphics[width=\columnwidth]{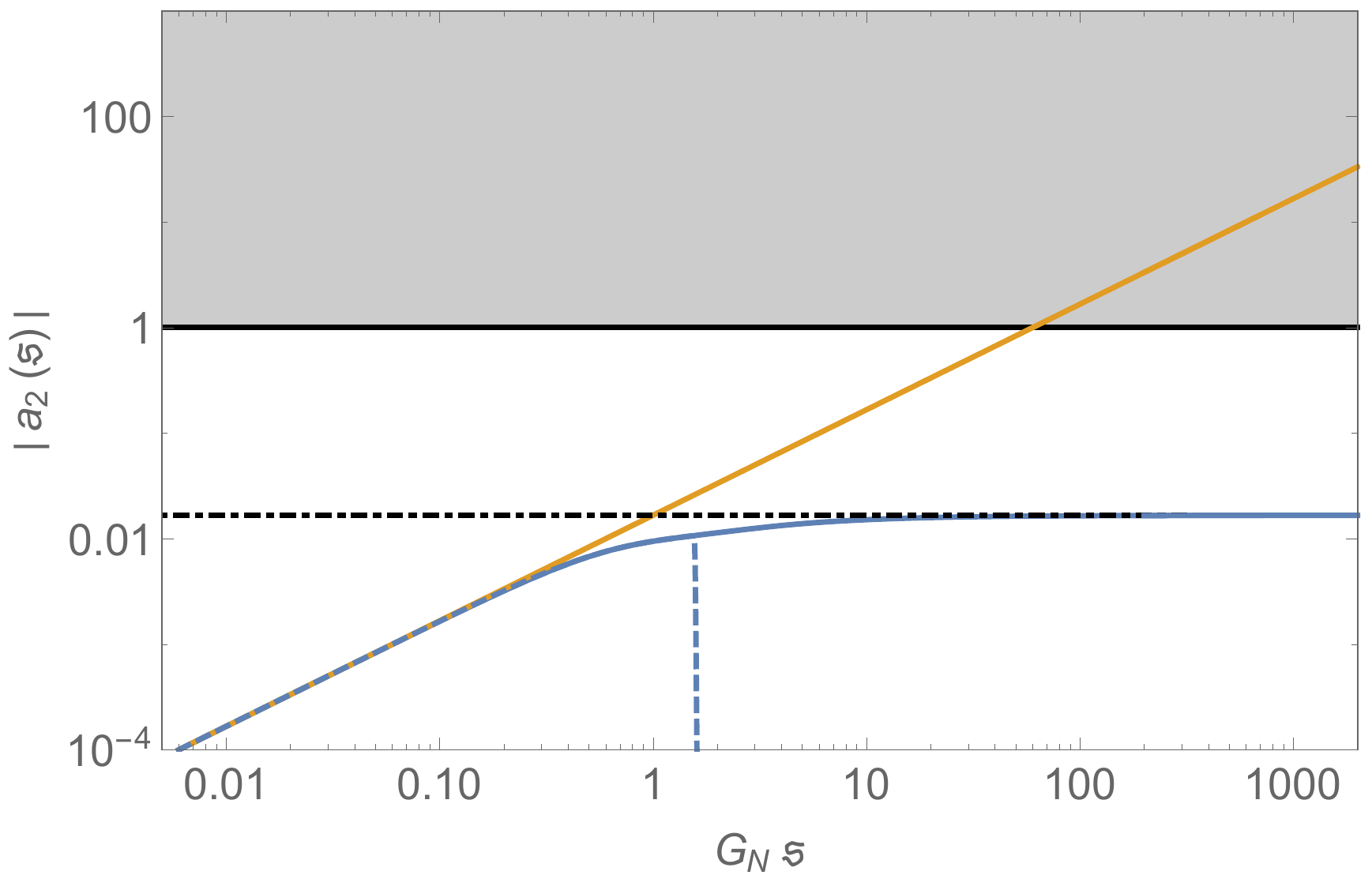}
	\caption{Illustration of the partial-wave amplitude $|a_2(\mans)|$ for our \modelname{} with $c_C = 1$ (blue solid line) and \GR{} (orange solid line). The horizontal lines show the  bound $|a_2(\mans)| = 1$ (solid black line) and the asymptotic value $\lim_{\mans \rightarrow \infty} |a_2| = 1/60$ (dash-dotted line). The dashed blue line indicates the amplitude from the truncated \modelname{} where the propagators \eqref{infiniteres} are expanded up to order $\mans^{250}$.}
	\label{fig:amplitude}
\end{figure}

Given the explicit form of the propagators \eqref{infiniteres} underlying our \modelname{}, it is straightforward to analyse the resulting partial-wave amplitudes \eqref{eq:partialwaves}.
Notably, $a_0(\mans)$ and $a_2(\mans)$ depend on $c_R$ and $c_C$ only and exhibit essentially the same qualitative behaviour.
This generic structure is exemplified for $a_2(\mans)$ in \autoref{fig:amplitude}. 
 For $\mans \lesssim c_C M_{\rm Pl}^2$, $a_2(\mans)$ agrees with the result from \GR{}.
 For $\mans \gtrsim  c_C M_{\rm Pl}^2$ the amplitudes are bounded and become scale-free as $\mans \to \infty$.
 Their maximum value is determined by the free parameters $c_R, c_C$:
 \be
 \lim_{\mans \rightarrow \infty} a_0(\mans) = \frac{1}{12c_R} \, , \quad 
 \lim_{\mans \rightarrow \infty} a_2(\mans) = - \frac{1}{60c_C}  \, . 
 \ee
Thus they are compatible with the bound \eqref{eq:ajbound} provided that $c_R \ge 1/12$ and $c_C \ge 1/60$.

Most importantly, the partial-wave amplitudes do not contain massive poles associated with new degrees of freedom or intermediate-state particles.
The flattening of the amplitude at trans-Planckian scales originates from infinite towers of \emph{massless} poles in the spin zero and spin two propagators located on the imaginary axis of the complex $\mans$-plane.
For sufficiently large $n$ the position of the poles follows a Regge-type behaviour, 
\be\label{Reggebehavior}
\ressym_{n,\pm}^{C} \simeq  \pm \frac{ \mathbf{i} \, \pi \, n }{c_C} M_{\rm Pl}^2   \, , \quad \ressym_{n,\pm}^{R} \simeq \pm \frac{\mathbf{i} \, \pi \, n }{c_R}M_{\rm Pl}^2 \, . 
\ee
The interplay of these poles leads to scattering amplitudes which are asymptotically scale-free and obey the conditions 1) -- 3) without introducing additional massive degrees of freedom. 
The role of the Regge behaviour \eqref{Reggebehavior} is thus manifestly different from the one in string theory, where the interplay of the infinite tower of higher-spin resonances ensures
the exponential decrease and causality of the amplitude \cite{DAppollonio:2015fly}.

\textit{Finite crossed amplitudes from matter form factors.}---
At this stage it is instructive to investigate the scattering amplitude related to $\phi\chi \rightarrow \phi\chi$-scattering which arises from the $\phi\phi \rightarrow \chi\chi$-amplitude by crossing symmetry $\mans \leftrightarrow \mant$.
For GR the result is
\be\label{crossedscatteringGR}
\mathcal A_\mant^{\text{GR}} = 8 \pi G_N  \frac{\mans \manu}{\mant} = - 8\pi G_N \frac{\mans(\mans+\mant)}{\mant} \, . 
\ee
For forward scattering, $\mans \rightarrow \infty$ at fixed $\mant$, this amplitude diverges quadratically, thus violating condition 2).
Inserting the gravitational propagators \eqref{IRMprop} leads to the replacement $\mant^{-1} \rightarrow G_C(\mant)$ and does not tame this divergence.
At this point the contribution of the form factor in the matter sector becomes crucial.
The on-shell vertex of the scalar self-interaction ({\small{SSI}}) gives the following contribution to the $\mans{}$-channel amplitude:
	\begin{equation}\label{matterff1}
\mathcal A_\mans^\text{SSI} = f_{\phi\chi}\left(\frac{\mans}{2},\frac{\mant}{2},\frac{\manu}{2},\frac{\manu}{2},\frac{\mant}{2},\frac{\mans}{2}\right) + \text{sym} \, ,
\end{equation}
where ``sym'' indicates the symmetrisation of the arguments as provided by functional variation of the action. Effectively, this can be parameterised by a function of three arguments,
$\mathcal A_\mans^\text{SSI} \equiv g(\mans|\mant,\manu)$,
which is symmetric under exchange of its last two arguments. Clearly different choices of $f$ can give rise to the same $g$ and thus the same amplitude, so in the following we will only define $g$. Specifically, we choose
\begin{equation}\label{matterpara}
g(a|x,y) = 4\pi G_N G_C(a) (x^2+y^2) f^{\text{int}}(a^2+x^2+y^2) \, .
\end{equation}
The function
\be
f^{\rm int}(x) =  \frac{c_t \, G_N^2 \, x \tanh[c_t G_N^2 \, x]}{1 + c_t \, G_N^2 \, x \tanh[c_t G_N^2 \, x]}
\ee
is manifestly invariant under crossing symmetry.
This factor ensures that the vertex contribution gives no essential contribution to the scattering amplitude at low energy so that \GR{} is recovered for $\mans \lesssim M_{\rm Pl}^2/c_t$.
The parameter $c_t$  sets the scale where the self-interaction starts contributing.

\begin{figure}
	\includegraphics[width=\columnwidth]{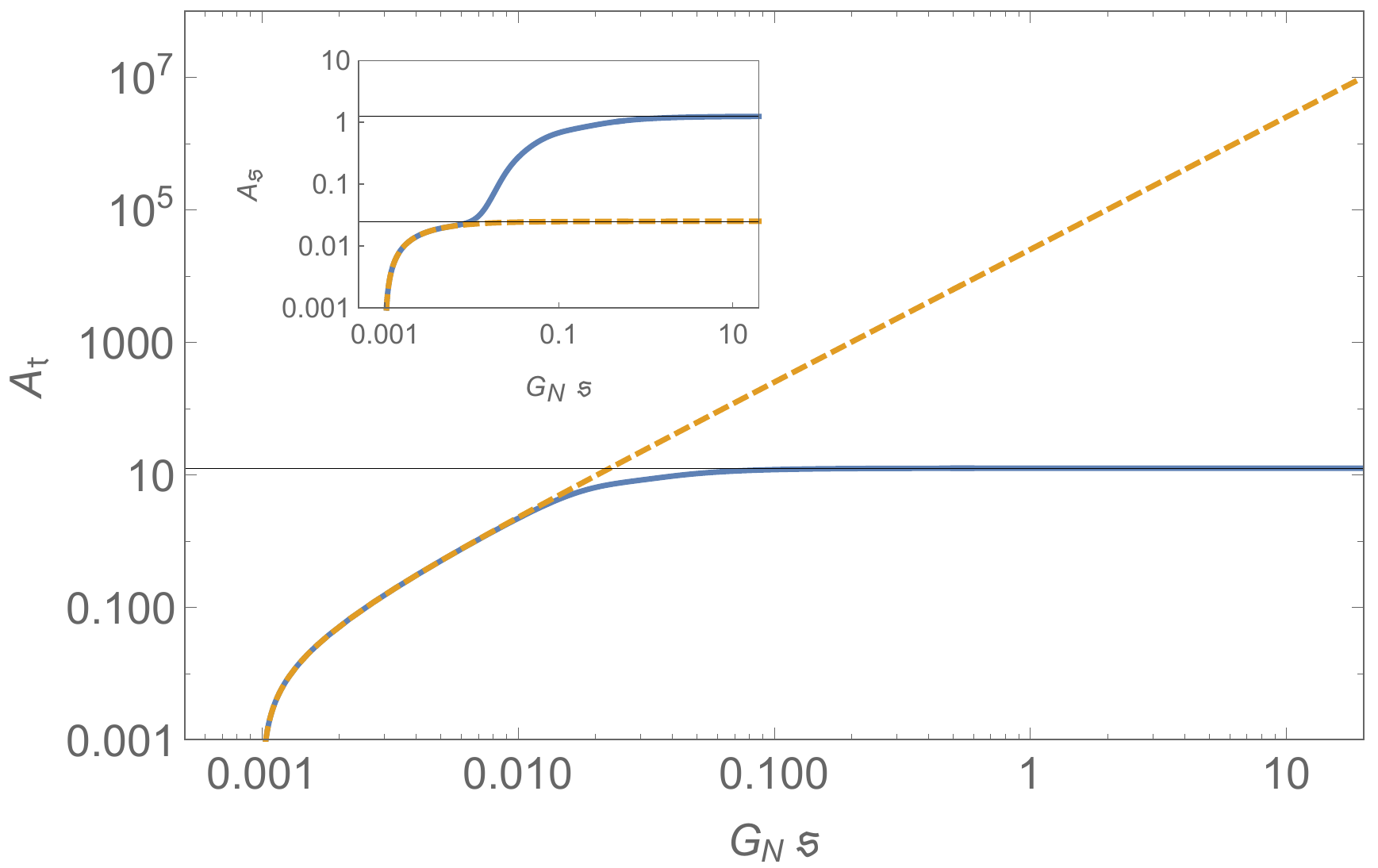}
	\caption{Illustration of the $\mant$-channel amplitude \eqref{amps} in the forward-scattering limit with $\mant = -0.001 M_{\rm Pl}^2$ held fixed with parameters $c_R = c_C = 10$  and $c_t = 10^3$  (solid line). The result obtained from \GR{} is given by the dashed line for comparison. The horizontal lines indicates the asymptotic value of the amplitude as $\mans \rightarrow \infty$. Inset: Illustration of the $\mans$-channel amplitude \eqref{amps} with (solid line) and without (dashed line) the contribution of the scalar self-interaction \eqref{matterpara}.}
	\label{fig:forwardamplitude}
\end{figure}

Including the self-interaction, the full scattering amplitudes of the \modelname{} are 
\begin{equation}\label{amps}
\begin{aligned}
 \mathcal A_\mans &= \frac{4\pi G_N}{3} \mans^2 \bigg[ G_R(\mans) -	\tfrac{\mans^2+6\mans\mant+6\mant^2}{\mans^2} G_C(\mans) \bigg] + g(\mans|\mant,\manu) \, , \\
 \mathcal A_\mant &= \frac{4\pi G_N}{3} \mant^2 \bigg[ G_R(\mant) -	\tfrac{6\mans^2+6\mans\mant+\mant^2}{\mant^2} G_C(\mant) \bigg] + g(\mant|\mans,\manu) \, .
\end{aligned}
\end{equation}
The forward-scattering amplitude entailed by \eqref{amps} is shown in \autoref{fig:forwardamplitude}.
Remarkably, the amplitude $\mathcal A_\mant$  becomes scale-free in the forward scattering limit. 
Thus the interplay of the graviton propagator with the self-interaction ensures that the model is compatible with condition 2) \emph{without violating any of the general properties of $\mathcal A_\mans$}.

The partial-wave decomposition of $\mathcal A_\mans$ shows that the self-interaction generates all partial-wave amplitudes $a_j(\mans)$ with $j$ even.
This accounts for the expectation that vertices in the quantum effective action should incorporate the contribution of ``ladder diagrams'' where the exchange of $n$ gravitons  generates a contribution to the partial-wave amplitude $a_{2n}(\mans)$.
Studying the high-energy asymptotics of the partial-wave amplitudes reveals
\be\label{spinres}
 \lim_{\mans \rightarrow \infty} a_j(\mans) = \left( \frac{1}{6c_C} + \frac{1}{12c_R} \right) \, \delta_{j,0} \, .
\ee
Thus the inclusion of the contact interaction softens the high-energy behaviour of $a_j(\mans)$. While this is not strictly required by condition 1), this fall-off is actually dictated by crossing symmetry in combination with condition 2).

%

\textit{Preservation of microcausality.}---
A characteristic feature of Lee-Wick models \cite{Lee:1969fy, Lee:1970iw}, coming with poles in the propagators located at complex momentum, is the violation of causality at time scales set by the imaginary part of the resonances \cite{Grinstein:2008bg, Anselmi:2018tmf, Donoghue:2019ecz}.
Following \cite{Grinstein:2008bg}, one finds that our amplitude preserves microcausality.
The taming of the amplitude \eqref{scats} is provided by a multiplicative factor which is regular for all (real) values of the momentum. As a result, the  frequency dependence of the matrix element relating in and out states is the one of a standard massless field theory where terms signalling the violation of microcausality are absent.
This analysis also reveals that the taming of the amplitude \eqref{scats} does not result from narrow Lee-Wick resonances (with formally vanishing decay time), but arises from the collective interplay of the poles characterising our construction.

\textit{Prediction for the location of the lowest resonance.}---
Combining the asymptotics \eqref{spinres} with the requirement \eqref{eq:ajbound} gives an upper bound on the energy scale where the existence of the imaginary poles must become visible.
Extremising \eqref{spinres} yields $c_C = c_R = 1/4$ so that
\be
|\ressym_{1,\pm}| \lesssim  3.4 \, M_{\rm Pl}^2 \, .
\ee

\textit{Quantum gravity requires non-perturbative physics.}---Notably, our construction allows to precisely pinpoint the non-perturbative physics making the model work.
For this purpose, it is instructive to perform a series expansion of the form factors \eqref{infiniteres} around $\Delta = 0$ which is equivalent to an expansion in $G_N$.
Terminating the expansion at a finite order and analysing the resulting pole structure shows  that the truncation reproduces the lowest-lying pole only.
All other poles provided by the truncation accumulate on a circle marking the finite radius of convergence of the expansion. This results in  the steep drop of $\cA_\mans$ at the radius of convergence, cf.\ the dashed line in \autoref{fig:amplitude}. 

\textit{Relation to Asymptotic Safety.}---
Our construction is closely related to the gravitational asymptotic safety program \cite{Weinberg:1976xy,Weinberg:1980gg}, where physical quantities like the scattering amplitudes considered in this work are expected to be finite.
The relations between the propagators and vertices should then be generated by an interacting fixed point of the theory's renormalisation group flow which ensures that the theory is scale-free at trans-Planckian energy.
A first principle computation deriving the pole structure underlying this work from the Wetterich equation for gravity \cite{Reuter:1996cp} then requires the full momentum dependence of the renormalised graviton propagator and low-order vertices.
The form factor program  \cite{Becker:2017tcx,Bosma:2019aiu,Knorr:2019atm} and the momentum dependence studied within the vertex expansion \cite{Christiansen:2014raa,Christiansen:2015rva,Meibohm:2015twa,Denz:2016qks,Christiansen:2017cxa,Eichhorn:2018ydy,Reichert:2020mja} constitute first steps in this direction.
Our results then provide an important proof of principle that the quantum effective action comes with sufficient freedom to accommodate the physics requirements of Asymptotic Safety.
Our results also indicate that, at the level of observables, the Asymptotic Safety construction requires intricate relations between different vertices in the high-energy limit to tame, \eg, the forward scattering amplitude.
Intriguingly, the results on momentum locality \cite{Christiansen:2015rva, Denz:2016qks} and effective universality \cite{Eichhorn:2018akn, Eichhorn:2018ydy, Meibohm:2015twa, Dona:2015tnf, Eichhorn:2017sok, Christiansen:2017cxa, Eichhorn:2018nda} suggest that such relations  are indeed realised by the underlying fixed point.

\textit{Conclusions.}---In this work we used the quantum effective action to study the gravity-mediated scattering of scalar fields. 
Introducing an infinite tower of massless (Lee-Wick type) poles with a formally instantaneous decay time leads to scattering amplitudes \emph{which are scale-free at trans-Planckian energy}.
This distinguishes our construction from string theory \cite{Green:1987sp} and infinite derivative gravity \cite{Biswas:2011ar,Talaganis:2016ovm, Talaganis:2014ida, Buoninfante:2018xiw,Buoninfante:2018mre, Modesto:2017sdr} where the amplitudes exhibit an exponential fall-off above a given threshold scale.\footnote{In spirit, our construction is similar to the weakly non-local (quasi-polynomial) gravitational theories introduced in \cite{Modesto:2014lga,Modesto:2015lna,Dona:2015tra}.}
Our amplitudes obey crossing symmetry by construction, are compatible with microcausality, and stay within unitarity bounds.
Our construction may serve as a benchmark for a broad range of quantum gravity programs.
In particular, it is relevant for the asymptotic safety program \cite{Percacci:2017fkn, Reuter:2019byg, Donoghue:2019clr, Bonanno:2020bil}, where the pole structure introduced in this work is a candidate for the physics generating the renormalisation group fixed point at the core of the program.

\bigskip

\acknowledgments

We would like to thank Wim Beenakker, Anupam Mazumdar, Alessia Platania, Les\l{}aw Rachwa\l{}, Martin Reuter and Melissa van Beekveld for interesting discussions. B.\ K.\ acknowledges support by Perimeter Institute for Theoretical Physics. Research at Perimeter Institute is supported in part by the Government of Canada through the Department of Innovation, Science and Economic Development Canada and by the Province of Ontario through the Ministry of Colleges and Universities. F.\ S.\  acknowledges financial support from  the Netherlands Organisation for Scientific Research (NWO) within the Foundation for Fundamental Research on Matter (FOM) grant 13VP12.

\bibliographystyle{apsrev4-1}
\bibliography{general_bib}

\begin{thebibliography}{73}%
\makeatletter
\providecommand \@ifxundefined [1]{%
 \@ifx{#1\undefined}
}%
\providecommand \@ifnum [1]{%
 \ifnum #1\expandafter \@firstoftwo
 \else \expandafter \@secondoftwo
 \fi
}%
\providecommand \@ifx [1]{%
 \ifx #1\expandafter \@firstoftwo
 \else \expandafter \@secondoftwo
 \fi
}%
\providecommand \natexlab [1]{#1}%
\providecommand \enquote  [1]{``#1''}%
\providecommand \bibnamefont  [1]{#1}%
\providecommand \bibfnamefont [1]{#1}%
\providecommand \citenamefont [1]{#1}%
\providecommand \href@noop [0]{\@secondoftwo}%
\providecommand \href [0]{\begingroup \@sanitize@url \@href}%
\providecommand \@href[1]{\@@startlink{#1}\@@href}%
\providecommand \@@href[1]{\endgroup#1\@@endlink}%
\providecommand \@sanitize@url [0]{\catcode `\\12\catcode `\$12\catcode
  `\&12\catcode `\#12\catcode `\^12\catcode `\_12\catcode `\%12\relax}%
\providecommand \@@startlink[1]{}%
\providecommand \@@endlink[0]{}%
\providecommand \url  [0]{\begingroup\@sanitize@url \@url }%
\providecommand \@url [1]{\endgroup\@href {#1}{\urlprefix }}%
\providecommand \urlprefix  [0]{URL }%
\providecommand \Eprint [0]{\href }%
\providecommand \doibase [0]{http://dx.doi.org/}%
\providecommand \selectlanguage [0]{\@gobble}%
\providecommand \bibinfo  [0]{\@secondoftwo}%
\providecommand \bibfield  [0]{\@secondoftwo}%
\providecommand \translation [1]{[#1]}%
\providecommand \BibitemOpen [0]{}%
\providecommand \bibitemStop [0]{}%
\providecommand \bibitemNoStop [0]{.\EOS\space}%
\providecommand \EOS [0]{\spacefactor3000\relax}%
\providecommand \BibitemShut  [1]{\csname bibitem#1\endcsname}%
\let\auto@bib@innerbib\@empty
\bibitem [{\citenamefont {'t~Hooft}\ and\ \citenamefont
  {Veltman}(1974)}]{'tHooft:1974bx}%
  \BibitemOpen
  \bibfield  {author} {\bibinfo {author} {\bibfnamefont {G.}~\bibnamefont
  {'t~Hooft}}\ and\ \bibinfo {author} {\bibfnamefont {M.~J.~G.}\ \bibnamefont
  {Veltman}},\ }\href@noop {} {\bibfield  {journal} {\bibinfo  {journal}
  {Annales Poincare Phys. Theor.}\ }\textbf {\bibinfo {volume} {A20}},\
  \bibinfo {pages} {69} (\bibinfo {year} {1974})}\BibitemShut {NoStop}%
\bibitem [{\citenamefont {Goroff}\ and\ \citenamefont
  {Sagnotti}(1985)}]{Goroff:1985sz}%
  \BibitemOpen
  \bibfield  {author} {\bibinfo {author} {\bibfnamefont {M.~H.}\ \bibnamefont
  {Goroff}}\ and\ \bibinfo {author} {\bibfnamefont {A.}~\bibnamefont
  {Sagnotti}},\ }\href {\doibase 10.1016/0370-2693(85)91470-4} {\bibfield
  {journal} {\bibinfo  {journal} {Phys. Lett.}\ }\textbf {\bibinfo {volume}
  {B160}},\ \bibinfo {pages} {81} (\bibinfo {year} {1985})}\BibitemShut
  {NoStop}%
\bibitem [{\citenamefont {Goroff}\ and\ \citenamefont
  {Sagnotti}(1986)}]{Goroff:1985th}%
  \BibitemOpen
  \bibfield  {author} {\bibinfo {author} {\bibfnamefont {M.~H.}\ \bibnamefont
  {Goroff}}\ and\ \bibinfo {author} {\bibfnamefont {A.}~\bibnamefont
  {Sagnotti}},\ }\href {\doibase 10.1016/0550-3213(86)90193-8} {\bibfield
  {journal} {\bibinfo  {journal} {Nucl. Phys.}\ }\textbf {\bibinfo {volume}
  {B266}},\ \bibinfo {pages} {709} (\bibinfo {year} {1986})}\BibitemShut
  {NoStop}%
\bibitem [{\citenamefont {Anber}\ and\ \citenamefont
  {Donoghue}(2012)}]{Anber:2011ut}%
  \BibitemOpen
  \bibfield  {author} {\bibinfo {author} {\bibfnamefont {M.~M.}\ \bibnamefont
  {Anber}}\ and\ \bibinfo {author} {\bibfnamefont {J.~F.}\ \bibnamefont
  {Donoghue}},\ }\href {\doibase 10.1103/PhysRevD.85.104016} {\bibfield
  {journal} {\bibinfo  {journal} {Phys. Rev. D}\ }\textbf {\bibinfo {volume}
  {85}},\ \bibinfo {pages} {104016} (\bibinfo {year} {2012})},\ \Eprint
  {http://arxiv.org/abs/1111.2875} {arXiv:1111.2875 [hep-th]} \BibitemShut
  {NoStop}%
\bibitem [{\citenamefont {Donoghue}(1994{\natexlab{a}})}]{Donoghue:1993eb}%
  \BibitemOpen
  \bibfield  {author} {\bibinfo {author} {\bibfnamefont {J.~F.}\ \bibnamefont
  {Donoghue}},\ }\href {\doibase 10.1103/PhysRevLett.72.2996} {\bibfield
  {journal} {\bibinfo  {journal} {Phys. Rev. Lett.}\ }\textbf {\bibinfo
  {volume} {72}},\ \bibinfo {pages} {2996} (\bibinfo {year}
  {1994}{\natexlab{a}})},\ \Eprint {http://arxiv.org/abs/gr-qc/9310024}
  {arXiv:gr-qc/9310024 [gr-qc]} \BibitemShut {NoStop}%
\bibitem [{\citenamefont {Donoghue}(1994{\natexlab{b}})}]{Donoghue:1994dn}%
  \BibitemOpen
  \bibfield  {author} {\bibinfo {author} {\bibfnamefont {J.~F.}\ \bibnamefont
  {Donoghue}},\ }\href {\doibase 10.1103/PhysRevD.50.3874} {\bibfield
  {journal} {\bibinfo  {journal} {Phys. Rev.}\ }\textbf {\bibinfo {volume}
  {D50}},\ \bibinfo {pages} {3874} (\bibinfo {year} {1994}{\natexlab{b}})},\
  \Eprint {http://arxiv.org/abs/gr-qc/9405057} {arXiv:gr-qc/9405057 [gr-qc]}
  \BibitemShut {NoStop}%
\bibitem [{\citenamefont {Aydemir}\ \emph {et~al.}(2012)\citenamefont
  {Aydemir}, \citenamefont {Anber},\ and\ \citenamefont
  {Donoghue}}]{Aydemir:2012nz}%
  \BibitemOpen
  \bibfield  {author} {\bibinfo {author} {\bibfnamefont {U.}~\bibnamefont
  {Aydemir}}, \bibinfo {author} {\bibfnamefont {M.~M.}\ \bibnamefont {Anber}},
  \ and\ \bibinfo {author} {\bibfnamefont {J.~F.}\ \bibnamefont {Donoghue}},\
  }\href {\doibase 10.1103/PhysRevD.86.014025} {\bibfield  {journal} {\bibinfo
  {journal} {Phys. Rev. D}\ }\textbf {\bibinfo {volume} {86}},\ \bibinfo
  {pages} {014025} (\bibinfo {year} {2012})},\ \Eprint
  {http://arxiv.org/abs/1203.5153} {arXiv:1203.5153 [hep-ph]} \BibitemShut
  {NoStop}%
\bibitem [{\citenamefont {Adams}\ \emph {et~al.}(2006)\citenamefont {Adams},
  \citenamefont {Arkani-Hamed}, \citenamefont {Dubovsky}, \citenamefont
  {Nicolis},\ and\ \citenamefont {Rattazzi}}]{Adams:2006sv}%
  \BibitemOpen
  \bibfield  {author} {\bibinfo {author} {\bibfnamefont {A.}~\bibnamefont
  {Adams}}, \bibinfo {author} {\bibfnamefont {N.}~\bibnamefont {Arkani-Hamed}},
  \bibinfo {author} {\bibfnamefont {S.}~\bibnamefont {Dubovsky}}, \bibinfo
  {author} {\bibfnamefont {A.}~\bibnamefont {Nicolis}}, \ and\ \bibinfo
  {author} {\bibfnamefont {R.}~\bibnamefont {Rattazzi}},\ }\href {\doibase
  10.1088/1126-6708/2006/10/014} {\bibfield  {journal} {\bibinfo  {journal}
  {JHEP}\ }\textbf {\bibinfo {volume} {10}},\ \bibinfo {pages} {014} (\bibinfo
  {year} {2006})},\ \Eprint {http://arxiv.org/abs/hep-th/0602178}
  {arXiv:hep-th/0602178} \BibitemShut {NoStop}%
\bibitem [{\citenamefont {Camanho}\ \emph {et~al.}(2016)\citenamefont
  {Camanho}, \citenamefont {Edelstein}, \citenamefont {Maldacena},\ and\
  \citenamefont {Zhiboedov}}]{Camanho:2014apa}%
  \BibitemOpen
  \bibfield  {author} {\bibinfo {author} {\bibfnamefont {X.~O.}\ \bibnamefont
  {Camanho}}, \bibinfo {author} {\bibfnamefont {J.~D.}\ \bibnamefont
  {Edelstein}}, \bibinfo {author} {\bibfnamefont {J.}~\bibnamefont
  {Maldacena}}, \ and\ \bibinfo {author} {\bibfnamefont {A.}~\bibnamefont
  {Zhiboedov}},\ }\href {\doibase 10.1007/JHEP02(2016)020} {\bibfield
  {journal} {\bibinfo  {journal} {JHEP}\ }\textbf {\bibinfo {volume} {02}},\
  \bibinfo {pages} {020} (\bibinfo {year} {2016})},\ \Eprint
  {http://arxiv.org/abs/1407.5597} {arXiv:1407.5597 [hep-th]} \BibitemShut
  {NoStop}%
\bibitem [{\citenamefont {Bellazzini}\ \emph {et~al.}(2016)\citenamefont
  {Bellazzini}, \citenamefont {Cheung},\ and\ \citenamefont
  {Remmen}}]{Bellazzini:2015cra}%
  \BibitemOpen
  \bibfield  {author} {\bibinfo {author} {\bibfnamefont {B.}~\bibnamefont
  {Bellazzini}}, \bibinfo {author} {\bibfnamefont {C.}~\bibnamefont {Cheung}},
  \ and\ \bibinfo {author} {\bibfnamefont {G.~N.}\ \bibnamefont {Remmen}},\
  }\href {\doibase 10.1103/PhysRevD.93.064076} {\bibfield  {journal} {\bibinfo
  {journal} {Phys. Rev. D}\ }\textbf {\bibinfo {volume} {93}},\ \bibinfo
  {pages} {064076} (\bibinfo {year} {2016})},\ \Eprint
  {http://arxiv.org/abs/1509.00851} {arXiv:1509.00851 [hep-th]} \BibitemShut
  {NoStop}%
\bibitem [{\citenamefont {Chandrasekaran}\ \emph {et~al.}(2018)\citenamefont
  {Chandrasekaran}, \citenamefont {Remmen},\ and\ \citenamefont
  {Shahbazi-Moghaddam}}]{Chandrasekaran:2018qmx}%
  \BibitemOpen
  \bibfield  {author} {\bibinfo {author} {\bibfnamefont {V.}~\bibnamefont
  {Chandrasekaran}}, \bibinfo {author} {\bibfnamefont {G.~N.}\ \bibnamefont
  {Remmen}}, \ and\ \bibinfo {author} {\bibfnamefont {A.}~\bibnamefont
  {Shahbazi-Moghaddam}},\ }\href {\doibase 10.1007/JHEP11(2018)015} {\bibfield
  {journal} {\bibinfo  {journal} {JHEP}\ }\textbf {\bibinfo {volume} {11}},\
  \bibinfo {pages} {015} (\bibinfo {year} {2018})},\ \Eprint
  {http://arxiv.org/abs/1804.03153} {arXiv:1804.03153 [hep-th]} \BibitemShut
  {NoStop}%
\bibitem [{\citenamefont {Stelle}(1977)}]{Stelle:1976gc}%
  \BibitemOpen
  \bibfield  {author} {\bibinfo {author} {\bibfnamefont {K.}~\bibnamefont
  {Stelle}},\ }\href {\doibase 10.1103/PhysRevD.16.953} {\bibfield  {journal}
  {\bibinfo  {journal} {Phys. Rev. D}\ }\textbf {\bibinfo {volume} {16}},\
  \bibinfo {pages} {953} (\bibinfo {year} {1977})}\BibitemShut {NoStop}%
\bibitem [{\citenamefont {Stelle}(1978)}]{Stelle:1977ry}%
  \BibitemOpen
  \bibfield  {author} {\bibinfo {author} {\bibfnamefont {K.~S.}\ \bibnamefont
  {Stelle}},\ }\href {\doibase 10.1007/BF00760427} {\bibfield  {journal}
  {\bibinfo  {journal} {Gen. Rel. Grav.}\ }\textbf {\bibinfo {volume} {9}},\
  \bibinfo {pages} {353} (\bibinfo {year} {1978})}\BibitemShut {NoStop}%
\bibitem [{\citenamefont {Anselmi}\ and\ \citenamefont
  {Marino}(2020)}]{Anselmi:2019nie}%
  \BibitemOpen
  \bibfield  {author} {\bibinfo {author} {\bibfnamefont {D.}~\bibnamefont
  {Anselmi}}\ and\ \bibinfo {author} {\bibfnamefont {A.}~\bibnamefont
  {Marino}},\ }\href {\doibase 10.1088/1361-6382/ab78d2} {\bibfield  {journal}
  {\bibinfo  {journal} {Class. Quant. Grav.}\ }\textbf {\bibinfo {volume}
  {37}},\ \bibinfo {pages} {095003} (\bibinfo {year} {2020})},\ \Eprint
  {http://arxiv.org/abs/1909.12873} {arXiv:1909.12873 [gr-qc]} \BibitemShut
  {NoStop}%
\bibitem [{\citenamefont {Donoghue}\ and\ \citenamefont
  {Menezes}(2019)}]{Donoghue:2019ecz}%
  \BibitemOpen
  \bibfield  {author} {\bibinfo {author} {\bibfnamefont {J.~F.}\ \bibnamefont
  {Donoghue}}\ and\ \bibinfo {author} {\bibfnamefont {G.}~\bibnamefont
  {Menezes}},\ }\href {\doibase 10.1103/PhysRevLett.123.171601} {\bibfield
  {journal} {\bibinfo  {journal} {Phys. Rev. Lett.}\ }\textbf {\bibinfo
  {volume} {123}},\ \bibinfo {pages} {171601} (\bibinfo {year} {2019})},\
  \Eprint {http://arxiv.org/abs/1908.04170} {arXiv:1908.04170 [hep-th]}
  \BibitemShut {NoStop}%
\bibitem [{\citenamefont {Tomboulis}(1977)}]{Tomboulis:1977jk}%
  \BibitemOpen
  \bibfield  {author} {\bibinfo {author} {\bibfnamefont {E.}~\bibnamefont
  {Tomboulis}},\ }\href {\doibase 10.1016/0370-2693(77)90678-5} {\bibfield
  {journal} {\bibinfo  {journal} {Phys. Lett. B}\ }\textbf {\bibinfo {volume}
  {70}},\ \bibinfo {pages} {361} (\bibinfo {year} {1977})}\BibitemShut
  {NoStop}%
\bibitem [{\citenamefont {Tomboulis}(1980)}]{Tomboulis:1980bs}%
  \BibitemOpen
  \bibfield  {author} {\bibinfo {author} {\bibfnamefont {E.}~\bibnamefont
  {Tomboulis}},\ }\href {\doibase 10.1016/0370-2693(80)90550-X} {\bibfield
  {journal} {\bibinfo  {journal} {Phys. Lett. B}\ }\textbf {\bibinfo {volume}
  {97}},\ \bibinfo {pages} {77} (\bibinfo {year} {1980})}\BibitemShut {NoStop}%
\bibitem [{\citenamefont {Anselmi}(2018)}]{Anselmi:2018kgz}%
  \BibitemOpen
  \bibfield  {author} {\bibinfo {author} {\bibfnamefont {D.}~\bibnamefont
  {Anselmi}},\ }\href {\doibase 10.1007/JHEP02(2018)141} {\bibfield  {journal}
  {\bibinfo  {journal} {JHEP}\ }\textbf {\bibinfo {volume} {02}},\ \bibinfo
  {pages} {141} (\bibinfo {year} {2018})},\ \Eprint
  {http://arxiv.org/abs/1801.00915} {arXiv:1801.00915 [hep-th]} \BibitemShut
  {NoStop}%
\bibitem [{\citenamefont {Anselmi}\ and\ \citenamefont
  {Piva}(2018)}]{Anselmi:2018tmf}%
  \BibitemOpen
  \bibfield  {author} {\bibinfo {author} {\bibfnamefont {D.}~\bibnamefont
  {Anselmi}}\ and\ \bibinfo {author} {\bibfnamefont {M.}~\bibnamefont {Piva}},\
  }\href {\doibase 10.1007/JHEP11(2018)021} {\bibfield  {journal} {\bibinfo
  {journal} {JHEP}\ }\textbf {\bibinfo {volume} {11}},\ \bibinfo {pages} {021}
  (\bibinfo {year} {2018})},\ \Eprint {http://arxiv.org/abs/1806.03605}
  {arXiv:1806.03605 [hep-th]} \BibitemShut {NoStop}%
\bibitem [{\citenamefont {Anselmi}(2019{\natexlab{a}})}]{Anselmi:2018bra}%
  \BibitemOpen
  \bibfield  {author} {\bibinfo {author} {\bibfnamefont {D.}~\bibnamefont
  {Anselmi}},\ }\href {\doibase 10.1088/1361-6382/ab04c8} {\bibfield  {journal}
  {\bibinfo  {journal} {Class. Quant. Grav.}\ }\textbf {\bibinfo {volume}
  {36}},\ \bibinfo {pages} {065010} (\bibinfo {year} {2019}{\natexlab{a}})},\
  \Eprint {http://arxiv.org/abs/1809.05037} {arXiv:1809.05037 [hep-th]}
  \BibitemShut {NoStop}%
\bibitem [{\citenamefont {Anselmi}(2020)}]{Anselmi:2020tqo}%
  \BibitemOpen
  \bibfield  {author} {\bibinfo {author} {\bibfnamefont {D.}~\bibnamefont
  {Anselmi}},\ }\href {\doibase 10.1007/JHEP03(2020)142} {\bibfield  {journal}
  {\bibinfo  {journal} {JHEP}\ }\textbf {\bibinfo {volume} {03}},\ \bibinfo
  {pages} {142} (\bibinfo {year} {2020})},\ \Eprint
  {http://arxiv.org/abs/2001.01942} {arXiv:2001.01942 [hep-th]} \BibitemShut
  {NoStop}%
\bibitem [{\citenamefont {Shapiro}(2015)}]{Shapiro:2015uxa}%
  \BibitemOpen
  \bibfield  {author} {\bibinfo {author} {\bibfnamefont {I.~L.}\ \bibnamefont
  {Shapiro}},\ }\href {\doibase 10.1016/j.physletb.2015.03.037} {\bibfield
  {journal} {\bibinfo  {journal} {Phys. Lett. B}\ }\textbf {\bibinfo {volume}
  {744}},\ \bibinfo {pages} {67} (\bibinfo {year} {2015})},\ \Eprint
  {http://arxiv.org/abs/1502.00106} {arXiv:1502.00106 [hep-th]} \BibitemShut
  {NoStop}%
\bibitem [{\citenamefont {de~O.Salles}\ and\ \citenamefont
  {Shapiro}(2018)}]{Salles:2018ccb}%
  \BibitemOpen
  \bibfield  {author} {\bibinfo {author} {\bibfnamefont {F.}~\bibnamefont
  {de~O.Salles}}\ and\ \bibinfo {author} {\bibfnamefont {I.~L.}\ \bibnamefont
  {Shapiro}},\ }\href {\doibase 10.3390/universe4090091} {\bibfield  {journal}
  {\bibinfo  {journal} {Universe}\ }\textbf {\bibinfo {volume} {4}},\ \bibinfo
  {pages} {91} (\bibinfo {year} {2018})},\ \Eprint
  {http://arxiv.org/abs/1808.09015} {arXiv:1808.09015 [gr-qc]} \BibitemShut
  {NoStop}%
\bibitem [{\citenamefont {Veneziano}(1968)}]{Veneziano:1968yb}%
  \BibitemOpen
  \bibfield  {author} {\bibinfo {author} {\bibfnamefont {G.}~\bibnamefont
  {Veneziano}},\ }\href {\doibase 10.1007/BF02824451} {\bibfield  {journal}
  {\bibinfo  {journal} {Nuovo Cim. A}\ }\textbf {\bibinfo {volume} {57}},\
  \bibinfo {pages} {190} (\bibinfo {year} {1968})}\BibitemShut {NoStop}%
\bibitem [{\citenamefont {Virasoro}(1969)}]{Virasoro:1969me}%
  \BibitemOpen
  \bibfield  {author} {\bibinfo {author} {\bibfnamefont {M.}~\bibnamefont
  {Virasoro}},\ }\href {\doibase 10.1103/PhysRev.177.2309} {\bibfield
  {journal} {\bibinfo  {journal} {Phys. Rev.}\ }\textbf {\bibinfo {volume}
  {177}},\ \bibinfo {pages} {2309} (\bibinfo {year} {1969})}\BibitemShut
  {NoStop}%
\bibitem [{\citenamefont {Shapiro}(1970)}]{Shapiro:1970gy}%
  \BibitemOpen
  \bibfield  {author} {\bibinfo {author} {\bibfnamefont {J.~A.}\ \bibnamefont
  {Shapiro}},\ }\href {\doibase 10.1016/0370-2693(70)90255-8} {\bibfield
  {journal} {\bibinfo  {journal} {Phys. Lett. B}\ }\textbf {\bibinfo {volume}
  {33}},\ \bibinfo {pages} {361} (\bibinfo {year} {1970})}\BibitemShut
  {NoStop}%
\bibitem [{\citenamefont {Alonso}\ and\ \citenamefont
  {Urbano}(2019)}]{Alonso:2019ptb}%
  \BibitemOpen
  \bibfield  {author} {\bibinfo {author} {\bibfnamefont {R.}~\bibnamefont
  {Alonso}}\ and\ \bibinfo {author} {\bibfnamefont {A.}~\bibnamefont
  {Urbano}},\ }\href {\doibase 10.1103/PhysRevD.100.095013} {\bibfield
  {journal} {\bibinfo  {journal} {Phys. Rev. D}\ }\textbf {\bibinfo {volume}
  {100}},\ \bibinfo {pages} {095013} (\bibinfo {year} {2019})},\ \Eprint
  {http://arxiv.org/abs/1906.11687} {arXiv:1906.11687 [hep-ph]} \BibitemShut
  {NoStop}%
\bibitem [{\citenamefont {Green}\ \emph {et~al.}(1988)\citenamefont {Green},
  \citenamefont {Schwarz},\ and\ \citenamefont {Witten}}]{Green:1987sp}%
  \BibitemOpen
  \bibfield  {author} {\bibinfo {author} {\bibfnamefont {M.~B.}\ \bibnamefont
  {Green}}, \bibinfo {author} {\bibfnamefont {J.}~\bibnamefont {Schwarz}}, \
  and\ \bibinfo {author} {\bibfnamefont {E.}~\bibnamefont {Witten}},\
  }\href@noop {} {\emph {\bibinfo {title} {{Superstring Theory, Vol. 1:
  Introduction}}}},\ Cambridge Monographs on Mathematical Physics\ (\bibinfo
  {year} {1988})\BibitemShut {NoStop}%
\bibitem [{\citenamefont {Douglas}(2003)}]{Douglas:2003um}%
  \BibitemOpen
  \bibfield  {author} {\bibinfo {author} {\bibfnamefont {M.~R.}\ \bibnamefont
  {Douglas}},\ }\href {\doibase 10.1088/1126-6708/2003/05/046} {\bibfield
  {journal} {\bibinfo  {journal} {JHEP}\ }\textbf {\bibinfo {volume} {05}},\
  \bibinfo {pages} {046} (\bibinfo {year} {2003})},\ \Eprint
  {http://arxiv.org/abs/hep-th/0303194} {arXiv:hep-th/0303194} \BibitemShut
  {NoStop}%
\bibitem [{\citenamefont {Taylor}\ and\ \citenamefont
  {Wang}(2015)}]{Taylor:2015xtz}%
  \BibitemOpen
  \bibfield  {author} {\bibinfo {author} {\bibfnamefont {W.}~\bibnamefont
  {Taylor}}\ and\ \bibinfo {author} {\bibfnamefont {Y.-N.}\ \bibnamefont
  {Wang}},\ }\href {\doibase 10.1007/JHEP12(2015)164} {\bibfield  {journal}
  {\bibinfo  {journal} {JHEP}\ }\textbf {\bibinfo {volume} {12}},\ \bibinfo
  {pages} {164} (\bibinfo {year} {2015})},\ \Eprint
  {http://arxiv.org/abs/1511.03209} {arXiv:1511.03209 [hep-th]} \BibitemShut
  {NoStop}%
\bibitem [{\citenamefont {Barvinsky}\ and\ \citenamefont
  {Vilkovisky}(1990)}]{Barvinsky:1990up}%
  \BibitemOpen
  \bibfield  {author} {\bibinfo {author} {\bibfnamefont {A.}~\bibnamefont
  {Barvinsky}}\ and\ \bibinfo {author} {\bibfnamefont {G.}~\bibnamefont
  {Vilkovisky}},\ }\href {\doibase 10.1016/0550-3213(90)90047-H} {\bibfield
  {journal} {\bibinfo  {journal} {Nucl. Phys. B}\ }\textbf {\bibinfo {volume}
  {333}},\ \bibinfo {pages} {471} (\bibinfo {year} {1990})}\BibitemShut
  {NoStop}%
\bibitem [{\citenamefont {Draper}\ \emph {et~al.}(2020)\citenamefont {Draper},
  \citenamefont {Knorr}, \citenamefont {Ripken},\ and\ \citenamefont
  {Saueressig}}]{Draper:2020knh}%
  \BibitemOpen
  \bibfield  {author} {\bibinfo {author} {\bibfnamefont {T.}~\bibnamefont
  {Draper}}, \bibinfo {author} {\bibfnamefont {B.}~\bibnamefont {Knorr}},
  \bibinfo {author} {\bibfnamefont {C.}~\bibnamefont {Ripken}}, \ and\ \bibinfo
  {author} {\bibfnamefont {F.}~\bibnamefont {Saueressig}},\ }\href@noop {} {\
  (\bibinfo {year} {2020})},\ \Eprint {http://arxiv.org/abs/2007.04396}
  {arXiv:2007.04396 [hep-th]} \BibitemShut {NoStop}%
\bibitem [{\citenamefont {Froissart}(2010)}]{Froissart:2010}%
  \BibitemOpen
  \bibfield  {author} {\bibinfo {author} {\bibfnamefont {M.}~\bibnamefont
  {Froissart}},\ }\href {\doibase 10.4249/scholarpedia.10353} {\bibfield
  {journal} {\bibinfo  {journal} {Scholarpedia}\ }\textbf {\bibinfo {volume}
  {5}},\ \bibinfo {pages} {10353} (\bibinfo {year} {2010})},\ \bibinfo {note}
  {revision \#91280}\BibitemShut {NoStop}%
\bibitem [{\citenamefont {Anselmi}(2019{\natexlab{b}})}]{Anselmi:2019xac}%
  \BibitemOpen
  \bibfield  {author} {\bibinfo {author} {\bibfnamefont {D.}~\bibnamefont
  {Anselmi}},\ }in\ \href@noop {} {\emph {\bibinfo {booktitle} {{Progress and
  Visions in Quantum Theory in View of Gravity}: {Bridging foundations of
  physics and mathematics}}}}\ (\bibinfo {year} {2019})\ \Eprint
  {http://arxiv.org/abs/1911.10343} {arXiv:1911.10343 [hep-th]} \BibitemShut
  {NoStop}%
\bibitem [{\citenamefont {Froissart}(1961)}]{Froissart:1961ux}%
  \BibitemOpen
  \bibfield  {author} {\bibinfo {author} {\bibfnamefont {M.}~\bibnamefont
  {Froissart}},\ }\href {\doibase 10.1103/PhysRev.123.1053} {\bibfield
  {journal} {\bibinfo  {journal} {Phys. Rev.}\ }\textbf {\bibinfo {volume}
  {123}},\ \bibinfo {pages} {1053} (\bibinfo {year} {1961})}\BibitemShut
  {NoStop}%
\bibitem [{\citenamefont {Cerulus}\ and\ \citenamefont
  {Martin}(1964)}]{Cerulus:1964cjb}%
  \BibitemOpen
  \bibfield  {author} {\bibinfo {author} {\bibfnamefont {F.~A.}\ \bibnamefont
  {Cerulus}}\ and\ \bibinfo {author} {\bibfnamefont {A.}~\bibnamefont
  {Martin}},\ }\href {\doibase 10.1016/0031-9163(64)90807-8} {\bibfield
  {journal} {\bibinfo  {journal} {Phys. Lett.}\ }\textbf {\bibinfo {volume}
  {8}},\ \bibinfo {pages} {80} (\bibinfo {year} {1964})}\BibitemShut {NoStop}%
\bibitem [{\citenamefont {Epstein}\ and\ \citenamefont
  {Martin}(2019)}]{Epstein:2019zdn}%
  \BibitemOpen
  \bibfield  {author} {\bibinfo {author} {\bibfnamefont {H.}~\bibnamefont
  {Epstein}}\ and\ \bibinfo {author} {\bibfnamefont {A.}~\bibnamefont
  {Martin}},\ }\href {\doibase 10.1103/PhysRevD.99.114025} {\bibfield
  {journal} {\bibinfo  {journal} {Phys. Rev. D}\ }\textbf {\bibinfo {volume}
  {99}},\ \bibinfo {pages} {114025} (\bibinfo {year} {2019})},\ \Eprint
  {http://arxiv.org/abs/1903.00953} {arXiv:1903.00953 [hep-th]} \BibitemShut
  {NoStop}%
\bibitem [{\citenamefont {Giddings}\ and\ \citenamefont
  {Porto}(2010)}]{Giddings:2009gj}%
  \BibitemOpen
  \bibfield  {author} {\bibinfo {author} {\bibfnamefont {S.~B.}\ \bibnamefont
  {Giddings}}\ and\ \bibinfo {author} {\bibfnamefont {R.~A.}\ \bibnamefont
  {Porto}},\ }\href {\doibase 10.1103/PhysRevD.81.025002} {\bibfield  {journal}
  {\bibinfo  {journal} {Phys. Rev. D}\ }\textbf {\bibinfo {volume} {81}},\
  \bibinfo {pages} {025002} (\bibinfo {year} {2010})},\ \Eprint
  {http://arxiv.org/abs/0908.0004} {arXiv:0908.0004 [hep-th]} \BibitemShut
  {NoStop}%
\bibitem [{\citenamefont {Dvali}\ \emph {et~al.}(2011)\citenamefont {Dvali},
  \citenamefont {Giudice}, \citenamefont {Gomez},\ and\ \citenamefont
  {Kehagias}}]{Dvali:2010jz}%
  \BibitemOpen
  \bibfield  {author} {\bibinfo {author} {\bibfnamefont {G.}~\bibnamefont
  {Dvali}}, \bibinfo {author} {\bibfnamefont {G.~F.}\ \bibnamefont {Giudice}},
  \bibinfo {author} {\bibfnamefont {C.}~\bibnamefont {Gomez}}, \ and\ \bibinfo
  {author} {\bibfnamefont {A.}~\bibnamefont {Kehagias}},\ }\href {\doibase
  10.1007/JHEP08(2011)108} {\bibfield  {journal} {\bibinfo  {journal} {JHEP}\
  }\textbf {\bibinfo {volume} {08}},\ \bibinfo {pages} {108} (\bibinfo {year}
  {2011})},\ \Eprint {http://arxiv.org/abs/1010.1415} {arXiv:1010.1415
  [hep-ph]} \BibitemShut {NoStop}%
\bibitem [{\citenamefont {D'Appollonio}\ \emph {et~al.}(2015)\citenamefont
  {D'Appollonio}, \citenamefont {Di~Vecchia}, \citenamefont {Russo},\ and\
  \citenamefont {Veneziano}}]{DAppollonio:2015fly}%
  \BibitemOpen
  \bibfield  {author} {\bibinfo {author} {\bibfnamefont {G.}~\bibnamefont
  {D'Appollonio}}, \bibinfo {author} {\bibfnamefont {P.}~\bibnamefont
  {Di~Vecchia}}, \bibinfo {author} {\bibfnamefont {R.}~\bibnamefont {Russo}}, \
  and\ \bibinfo {author} {\bibfnamefont {G.}~\bibnamefont {Veneziano}},\ }\href
  {\doibase 10.1007/JHEP05(2015)144} {\bibfield  {journal} {\bibinfo  {journal}
  {JHEP}\ }\textbf {\bibinfo {volume} {05}},\ \bibinfo {pages} {144} (\bibinfo
  {year} {2015})},\ \Eprint {http://arxiv.org/abs/1502.01254} {arXiv:1502.01254
  [hep-th]} \BibitemShut {NoStop}%
\bibitem [{\citenamefont {Lee}\ and\ \citenamefont {Wick}(1969)}]{Lee:1969fy}%
  \BibitemOpen
  \bibfield  {author} {\bibinfo {author} {\bibfnamefont {T.}~\bibnamefont
  {Lee}}\ and\ \bibinfo {author} {\bibfnamefont {G.}~\bibnamefont {Wick}},\
  }\href {\doibase 10.1016/0550-3213(69)90098-4} {\bibfield  {journal}
  {\bibinfo  {journal} {Nucl. Phys. B}\ }\textbf {\bibinfo {volume} {9}},\
  \bibinfo {pages} {209} (\bibinfo {year} {1969})}\BibitemShut {NoStop}%
\bibitem [{\citenamefont {Lee}\ and\ \citenamefont {Wick}(1970)}]{Lee:1970iw}%
  \BibitemOpen
  \bibfield  {author} {\bibinfo {author} {\bibfnamefont {T.}~\bibnamefont
  {Lee}}\ and\ \bibinfo {author} {\bibfnamefont {G.}~\bibnamefont {Wick}},\
  }\href {\doibase 10.1103/PhysRevD.2.1033} {\bibfield  {journal} {\bibinfo
  {journal} {Phys. Rev. D}\ }\textbf {\bibinfo {volume} {2}},\ \bibinfo {pages}
  {1033} (\bibinfo {year} {1970})}\BibitemShut {NoStop}%
\bibitem [{\citenamefont {Grinstein}\ \emph {et~al.}(2009)\citenamefont
  {Grinstein}, \citenamefont {O'Connell},\ and\ \citenamefont
  {Wise}}]{Grinstein:2008bg}%
  \BibitemOpen
  \bibfield  {author} {\bibinfo {author} {\bibfnamefont {B.}~\bibnamefont
  {Grinstein}}, \bibinfo {author} {\bibfnamefont {D.}~\bibnamefont
  {O'Connell}}, \ and\ \bibinfo {author} {\bibfnamefont {M.~B.}\ \bibnamefont
  {Wise}},\ }\href {\doibase 10.1103/PhysRevD.79.105019} {\bibfield  {journal}
  {\bibinfo  {journal} {Phys. Rev. D}\ }\textbf {\bibinfo {volume} {79}},\
  \bibinfo {pages} {105019} (\bibinfo {year} {2009})},\ \Eprint
  {http://arxiv.org/abs/0805.2156} {arXiv:0805.2156 [hep-th]} \BibitemShut
  {NoStop}%
\bibitem [{\citenamefont {Weinberg}(1976)}]{Weinberg:1976xy}%
  \BibitemOpen
  \bibfield  {author} {\bibinfo {author} {\bibfnamefont {S.}~\bibnamefont
  {Weinberg}},\ }in\ \href@noop {} {\emph {\bibinfo {booktitle} {{Erice
  Subnucl.Phys.1976:1}}}}\ (\bibinfo {year} {1976})\ p.~\bibinfo {pages}
  {1}\BibitemShut {NoStop}%
\bibitem [{\citenamefont {Weinberg}(1979)}]{Weinberg:1980gg}%
  \BibitemOpen
  \bibfield  {author} {\bibinfo {author} {\bibfnamefont {S.}~\bibnamefont
  {Weinberg}},\ }\href@noop {} {\bibfield  {journal} {\bibinfo  {journal}
  {General Relativity: An Einstein centenary survey, Eds. Hawking, S.W.,
  Israel, W; Cambridge University Press}\ ,\ \bibinfo {pages} {790}} (\bibinfo
  {year} {1979})}\BibitemShut {NoStop}%
\bibitem [{\citenamefont {Reuter}(1998)}]{Reuter:1996cp}%
  \BibitemOpen
  \bibfield  {author} {\bibinfo {author} {\bibfnamefont {M.}~\bibnamefont
  {Reuter}},\ }\href {\doibase 10.1103/PhysRevD.57.971} {\bibfield  {journal}
  {\bibinfo  {journal} {Phys.Rev.}\ }\textbf {\bibinfo {volume} {D57}},\
  \bibinfo {pages} {971} (\bibinfo {year} {1998})},\ \Eprint
  {http://arxiv.org/abs/hep-th/9605030} {arXiv:hep-th/9605030 [hep-th]}
  \BibitemShut {NoStop}%
\bibitem [{\citenamefont {Becker}\ \emph {et~al.}(2017)\citenamefont {Becker},
  \citenamefont {Ripken},\ and\ \citenamefont {Saueressig}}]{Becker:2017tcx}%
  \BibitemOpen
  \bibfield  {author} {\bibinfo {author} {\bibfnamefont {D.}~\bibnamefont
  {Becker}}, \bibinfo {author} {\bibfnamefont {C.}~\bibnamefont {Ripken}}, \
  and\ \bibinfo {author} {\bibfnamefont {F.}~\bibnamefont {Saueressig}},\
  }\href {\doibase 10.1007/JHEP12(2017)121} {\bibfield  {journal} {\bibinfo
  {journal} {JHEP}\ }\textbf {\bibinfo {volume} {12}},\ \bibinfo {pages} {121}
  (\bibinfo {year} {2017})},\ \Eprint {http://arxiv.org/abs/1709.09098}
  {arXiv:1709.09098 [hep-th]} \BibitemShut {NoStop}%
\bibitem [{\citenamefont {Bosma}\ \emph {et~al.}(2019)\citenamefont {Bosma},
  \citenamefont {Knorr},\ and\ \citenamefont {Saueressig}}]{Bosma:2019aiu}%
  \BibitemOpen
  \bibfield  {author} {\bibinfo {author} {\bibfnamefont {L.}~\bibnamefont
  {Bosma}}, \bibinfo {author} {\bibfnamefont {B.}~\bibnamefont {Knorr}}, \ and\
  \bibinfo {author} {\bibfnamefont {F.}~\bibnamefont {Saueressig}},\ }\href
  {\doibase 10.1103/PhysRevLett.123.101301} {\bibfield  {journal} {\bibinfo
  {journal} {Phys. Rev. Lett.}\ }\textbf {\bibinfo {volume} {123}},\ \bibinfo
  {pages} {101301} (\bibinfo {year} {2019})},\ \Eprint
  {http://arxiv.org/abs/1904.04845} {arXiv:1904.04845 [hep-th]} \BibitemShut
  {NoStop}%
\bibitem [{\citenamefont {Knorr}\ \emph {et~al.}(2019)\citenamefont {Knorr},
  \citenamefont {Ripken},\ and\ \citenamefont {Saueressig}}]{Knorr:2019atm}%
  \BibitemOpen
  \bibfield  {author} {\bibinfo {author} {\bibfnamefont {B.}~\bibnamefont
  {Knorr}}, \bibinfo {author} {\bibfnamefont {C.}~\bibnamefont {Ripken}}, \
  and\ \bibinfo {author} {\bibfnamefont {F.}~\bibnamefont {Saueressig}},\
  }\href {\doibase 10.1088/1361-6382/ab4a53} {\bibfield  {journal} {\bibinfo
  {journal} {Class. Quant. Grav.}\ }\textbf {\bibinfo {volume} {36}},\ \bibinfo
  {pages} {234001} (\bibinfo {year} {2019})},\ \Eprint
  {http://arxiv.org/abs/1907.02903} {arXiv:1907.02903 [hep-th]} \BibitemShut
  {NoStop}%
\bibitem [{\citenamefont {Christiansen}\ \emph {et~al.}(2016)\citenamefont
  {Christiansen}, \citenamefont {Knorr}, \citenamefont {Pawlowski},\ and\
  \citenamefont {Rodigast}}]{Christiansen:2014raa}%
  \BibitemOpen
  \bibfield  {author} {\bibinfo {author} {\bibfnamefont {N.}~\bibnamefont
  {Christiansen}}, \bibinfo {author} {\bibfnamefont {B.}~\bibnamefont {Knorr}},
  \bibinfo {author} {\bibfnamefont {J.~M.}\ \bibnamefont {Pawlowski}}, \ and\
  \bibinfo {author} {\bibfnamefont {A.}~\bibnamefont {Rodigast}},\ }\href
  {\doibase 10.1103/PhysRevD.93.044036} {\bibfield  {journal} {\bibinfo
  {journal} {Phys. Rev.}\ }\textbf {\bibinfo {volume} {D93}},\ \bibinfo {pages}
  {044036} (\bibinfo {year} {2016})},\ \Eprint {http://arxiv.org/abs/1403.1232}
  {arXiv:1403.1232 [hep-th]} \BibitemShut {NoStop}%
\bibitem [{\citenamefont {Christiansen}\ \emph {et~al.}(2015)\citenamefont
  {Christiansen}, \citenamefont {Knorr}, \citenamefont {Meibohm}, \citenamefont
  {Pawlowski},\ and\ \citenamefont {Reichert}}]{Christiansen:2015rva}%
  \BibitemOpen
  \bibfield  {author} {\bibinfo {author} {\bibfnamefont {N.}~\bibnamefont
  {Christiansen}}, \bibinfo {author} {\bibfnamefont {B.}~\bibnamefont {Knorr}},
  \bibinfo {author} {\bibfnamefont {J.}~\bibnamefont {Meibohm}}, \bibinfo
  {author} {\bibfnamefont {J.~M.}\ \bibnamefont {Pawlowski}}, \ and\ \bibinfo
  {author} {\bibfnamefont {M.}~\bibnamefont {Reichert}},\ }\href {\doibase
  10.1103/PhysRevD.92.121501} {\bibfield  {journal} {\bibinfo  {journal} {Phys.
  Rev.}\ }\textbf {\bibinfo {volume} {D92}},\ \bibinfo {pages} {121501}
  (\bibinfo {year} {2015})},\ \Eprint {http://arxiv.org/abs/1506.07016}
  {arXiv:1506.07016 [hep-th]} \BibitemShut {NoStop}%
\bibitem [{\citenamefont {Meibohm}\ \emph {et~al.}(2016)\citenamefont
  {Meibohm}, \citenamefont {Pawlowski},\ and\ \citenamefont
  {Reichert}}]{Meibohm:2015twa}%
  \BibitemOpen
  \bibfield  {author} {\bibinfo {author} {\bibfnamefont {J.}~\bibnamefont
  {Meibohm}}, \bibinfo {author} {\bibfnamefont {J.~M.}\ \bibnamefont
  {Pawlowski}}, \ and\ \bibinfo {author} {\bibfnamefont {M.}~\bibnamefont
  {Reichert}},\ }\href {\doibase 10.1103/PhysRevD.93.084035} {\bibfield
  {journal} {\bibinfo  {journal} {Phys. Rev.}\ }\textbf {\bibinfo {volume}
  {D93}},\ \bibinfo {pages} {084035} (\bibinfo {year} {2016})},\ \Eprint
  {http://arxiv.org/abs/1510.07018} {arXiv:1510.07018 [hep-th]} \BibitemShut
  {NoStop}%
\bibitem [{\citenamefont {Denz}\ \emph {et~al.}(2018)\citenamefont {Denz},
  \citenamefont {Pawlowski},\ and\ \citenamefont {Reichert}}]{Denz:2016qks}%
  \BibitemOpen
  \bibfield  {author} {\bibinfo {author} {\bibfnamefont {T.}~\bibnamefont
  {Denz}}, \bibinfo {author} {\bibfnamefont {J.~M.}\ \bibnamefont {Pawlowski}},
  \ and\ \bibinfo {author} {\bibfnamefont {M.}~\bibnamefont {Reichert}},\
  }\href {\doibase 10.1140/epjc/s10052-018-5806-0} {\bibfield  {journal}
  {\bibinfo  {journal} {Eur. Phys. J.}\ }\textbf {\bibinfo {volume} {C78}},\
  \bibinfo {pages} {336} (\bibinfo {year} {2018})},\ \Eprint
  {http://arxiv.org/abs/1612.07315} {arXiv:1612.07315 [hep-th]} \BibitemShut
  {NoStop}%
\bibitem [{\citenamefont {Christiansen}\ \emph {et~al.}(2018)\citenamefont
  {Christiansen}, \citenamefont {Litim}, \citenamefont {Pawlowski},\ and\
  \citenamefont {Reichert}}]{Christiansen:2017cxa}%
  \BibitemOpen
  \bibfield  {author} {\bibinfo {author} {\bibfnamefont {N.}~\bibnamefont
  {Christiansen}}, \bibinfo {author} {\bibfnamefont {D.~F.}\ \bibnamefont
  {Litim}}, \bibinfo {author} {\bibfnamefont {J.~M.}\ \bibnamefont
  {Pawlowski}}, \ and\ \bibinfo {author} {\bibfnamefont {M.}~\bibnamefont
  {Reichert}},\ }\href {\doibase 10.1103/PhysRevD.97.106012} {\bibfield
  {journal} {\bibinfo  {journal} {Phys. Rev.}\ }\textbf {\bibinfo {volume}
  {D97}},\ \bibinfo {pages} {106012} (\bibinfo {year} {2018})},\ \Eprint
  {http://arxiv.org/abs/1710.04669} {arXiv:1710.04669 [hep-th]} \BibitemShut
  {NoStop}%
\bibitem [{\citenamefont {Eichhorn}\ \emph
  {et~al.}(2019{\natexlab{a}})\citenamefont {Eichhorn}, \citenamefont
  {Lippoldt}, \citenamefont {Pawlowski}, \citenamefont {Reichert},\ and\
  \citenamefont {Schiffer}}]{Eichhorn:2018ydy}%
  \BibitemOpen
  \bibfield  {author} {\bibinfo {author} {\bibfnamefont {A.}~\bibnamefont
  {Eichhorn}}, \bibinfo {author} {\bibfnamefont {S.}~\bibnamefont {Lippoldt}},
  \bibinfo {author} {\bibfnamefont {J.~M.}\ \bibnamefont {Pawlowski}}, \bibinfo
  {author} {\bibfnamefont {M.}~\bibnamefont {Reichert}}, \ and\ \bibinfo
  {author} {\bibfnamefont {M.}~\bibnamefont {Schiffer}},\ }\href {\doibase
  10.1016/j.physletb.2019.01.071} {\bibfield  {journal} {\bibinfo  {journal}
  {Phys. Lett.}\ }\textbf {\bibinfo {volume} {B792}},\ \bibinfo {pages} {310}
  (\bibinfo {year} {2019}{\natexlab{a}})},\ \Eprint
  {http://arxiv.org/abs/1810.02828} {arXiv:1810.02828 [hep-th]} \BibitemShut
  {NoStop}%
\bibitem [{\citenamefont {Reichert}(2020)}]{Reichert:2020mja}%
  \BibitemOpen
  \bibfield  {author} {\bibinfo {author} {\bibfnamefont {M.}~\bibnamefont
  {Reichert}},\ }\href {\doibase 10.22323/1.384.0005} {\bibfield  {journal}
  {\bibinfo  {journal} {PoS}\ }\textbf {\bibinfo {volume} {Modave2019}},\
  \bibinfo {pages} {005} (\bibinfo {year} {2020})}\BibitemShut {NoStop}%
\bibitem [{\citenamefont {Eichhorn}\ \emph
  {et~al.}(2018{\natexlab{a}})\citenamefont {Eichhorn}, \citenamefont {Labus},
  \citenamefont {Pawlowski},\ and\ \citenamefont
  {Reichert}}]{Eichhorn:2018akn}%
  \BibitemOpen
  \bibfield  {author} {\bibinfo {author} {\bibfnamefont {A.}~\bibnamefont
  {Eichhorn}}, \bibinfo {author} {\bibfnamefont {P.}~\bibnamefont {Labus}},
  \bibinfo {author} {\bibfnamefont {J.~M.}\ \bibnamefont {Pawlowski}}, \ and\
  \bibinfo {author} {\bibfnamefont {M.}~\bibnamefont {Reichert}},\ }\href
  {\doibase 10.21468/SciPostPhys.5.4.031} {\bibfield  {journal} {\bibinfo
  {journal} {SciPost Phys.}\ }\textbf {\bibinfo {volume} {5}},\ \bibinfo
  {pages} {031} (\bibinfo {year} {2018}{\natexlab{a}})},\ \Eprint
  {http://arxiv.org/abs/1804.00012} {arXiv:1804.00012 [hep-th]} \BibitemShut
  {NoStop}%
\bibitem [{\citenamefont {Donà}\ \emph {et~al.}(2016)\citenamefont {Donà},
  \citenamefont {Eichhorn}, \citenamefont {Labus},\ and\ \citenamefont
  {Percacci}}]{Dona:2015tnf}%
  \BibitemOpen
  \bibfield  {author} {\bibinfo {author} {\bibfnamefont {P.}~\bibnamefont
  {Donà}}, \bibinfo {author} {\bibfnamefont {A.}~\bibnamefont {Eichhorn}},
  \bibinfo {author} {\bibfnamefont {P.}~\bibnamefont {Labus}}, \ and\ \bibinfo
  {author} {\bibfnamefont {R.}~\bibnamefont {Percacci}},\ }\href {\doibase
  10.1103/PhysRevD.93.129904, 10.1103/PhysRevD.93.044049} {\bibfield  {journal}
  {\bibinfo  {journal} {Phys. Rev.}\ }\textbf {\bibinfo {volume} {D93}},\
  \bibinfo {pages} {044049} (\bibinfo {year} {2016})},\ \bibinfo {note}
  {[Erratum: Phys. Rev.D93,no.12,129904(2016)]},\ \Eprint
  {http://arxiv.org/abs/1512.01589} {arXiv:1512.01589 [gr-qc]} \BibitemShut
  {NoStop}%
\bibitem [{\citenamefont {Eichhorn}\ \emph
  {et~al.}(2018{\natexlab{b}})\citenamefont {Eichhorn}, \citenamefont
  {Lippoldt},\ and\ \citenamefont {Skrinjar}}]{Eichhorn:2017sok}%
  \BibitemOpen
  \bibfield  {author} {\bibinfo {author} {\bibfnamefont {A.}~\bibnamefont
  {Eichhorn}}, \bibinfo {author} {\bibfnamefont {S.}~\bibnamefont {Lippoldt}},
  \ and\ \bibinfo {author} {\bibfnamefont {V.}~\bibnamefont {Skrinjar}},\
  }\href {\doibase 10.1103/PhysRevD.97.026002} {\bibfield  {journal} {\bibinfo
  {journal} {Phys. Rev. D}\ }\textbf {\bibinfo {volume} {97}},\ \bibinfo
  {pages} {026002} (\bibinfo {year} {2018}{\natexlab{b}})},\ \Eprint
  {http://arxiv.org/abs/1710.03005} {arXiv:1710.03005 [hep-th]} \BibitemShut
  {NoStop}%
\bibitem [{\citenamefont {Eichhorn}\ \emph
  {et~al.}(2019{\natexlab{b}})\citenamefont {Eichhorn}, \citenamefont
  {Lippoldt},\ and\ \citenamefont {Schiffer}}]{Eichhorn:2018nda}%
  \BibitemOpen
  \bibfield  {author} {\bibinfo {author} {\bibfnamefont {A.}~\bibnamefont
  {Eichhorn}}, \bibinfo {author} {\bibfnamefont {S.}~\bibnamefont {Lippoldt}},
  \ and\ \bibinfo {author} {\bibfnamefont {M.}~\bibnamefont {Schiffer}},\
  }\href {\doibase 10.1103/PhysRevD.99.086002} {\bibfield  {journal} {\bibinfo
  {journal} {Phys. Rev.}\ }\textbf {\bibinfo {volume} {D99}},\ \bibinfo {pages}
  {086002} (\bibinfo {year} {2019}{\natexlab{b}})},\ \Eprint
  {http://arxiv.org/abs/1812.08782} {arXiv:1812.08782 [hep-th]} \BibitemShut
  {NoStop}%
\bibitem [{\citenamefont {Biswas}\ \emph {et~al.}(2012)\citenamefont {Biswas},
  \citenamefont {Gerwick}, \citenamefont {Koivisto},\ and\ \citenamefont
  {Mazumdar}}]{Biswas:2011ar}%
  \BibitemOpen
  \bibfield  {author} {\bibinfo {author} {\bibfnamefont {T.}~\bibnamefont
  {Biswas}}, \bibinfo {author} {\bibfnamefont {E.}~\bibnamefont {Gerwick}},
  \bibinfo {author} {\bibfnamefont {T.}~\bibnamefont {Koivisto}}, \ and\
  \bibinfo {author} {\bibfnamefont {A.}~\bibnamefont {Mazumdar}},\ }\href
  {\doibase 10.1103/PhysRevLett.108.031101} {\bibfield  {journal} {\bibinfo
  {journal} {Phys. Rev. Lett.}\ }\textbf {\bibinfo {volume} {108}},\ \bibinfo
  {pages} {031101} (\bibinfo {year} {2012})},\ \Eprint
  {http://arxiv.org/abs/1110.5249} {arXiv:1110.5249 [gr-qc]} \BibitemShut
  {NoStop}%
\bibitem [{\citenamefont {Talaganis}\ and\ \citenamefont
  {Mazumdar}(2016)}]{Talaganis:2016ovm}%
  \BibitemOpen
  \bibfield  {author} {\bibinfo {author} {\bibfnamefont {S.}~\bibnamefont
  {Talaganis}}\ and\ \bibinfo {author} {\bibfnamefont {A.}~\bibnamefont
  {Mazumdar}},\ }\href {\doibase 10.1088/0264-9381/33/14/145005} {\bibfield
  {journal} {\bibinfo  {journal} {Class. Quant. Grav.}\ }\textbf {\bibinfo
  {volume} {33}},\ \bibinfo {pages} {145005} (\bibinfo {year} {2016})},\
  \Eprint {http://arxiv.org/abs/1603.03440} {arXiv:1603.03440 [hep-th]}
  \BibitemShut {NoStop}%
\bibitem [{\citenamefont {Talaganis}\ \emph {et~al.}(2015)\citenamefont
  {Talaganis}, \citenamefont {Biswas},\ and\ \citenamefont
  {Mazumdar}}]{Talaganis:2014ida}%
  \BibitemOpen
  \bibfield  {author} {\bibinfo {author} {\bibfnamefont {S.}~\bibnamefont
  {Talaganis}}, \bibinfo {author} {\bibfnamefont {T.}~\bibnamefont {Biswas}}, \
  and\ \bibinfo {author} {\bibfnamefont {A.}~\bibnamefont {Mazumdar}},\ }\href
  {\doibase 10.1088/0264-9381/32/21/215017} {\bibfield  {journal} {\bibinfo
  {journal} {Class. Quant. Grav.}\ }\textbf {\bibinfo {volume} {32}},\ \bibinfo
  {pages} {215017} (\bibinfo {year} {2015})},\ \Eprint
  {http://arxiv.org/abs/1412.3467} {arXiv:1412.3467 [hep-th]} \BibitemShut
  {NoStop}%
\bibitem [{\citenamefont {Buoninfante}\ \emph {et~al.}(2018)\citenamefont
  {Buoninfante}, \citenamefont {Koshelev}, \citenamefont {Lambiase},\ and\
  \citenamefont {Mazumdar}}]{Buoninfante:2018xiw}%
  \BibitemOpen
  \bibfield  {author} {\bibinfo {author} {\bibfnamefont {L.}~\bibnamefont
  {Buoninfante}}, \bibinfo {author} {\bibfnamefont {A.~S.}\ \bibnamefont
  {Koshelev}}, \bibinfo {author} {\bibfnamefont {G.}~\bibnamefont {Lambiase}},
  \ and\ \bibinfo {author} {\bibfnamefont {A.}~\bibnamefont {Mazumdar}},\
  }\href {\doibase 10.1088/1475-7516/2018/09/034} {\bibfield  {journal}
  {\bibinfo  {journal} {JCAP}\ }\textbf {\bibinfo {volume} {09}},\ \bibinfo
  {pages} {034} (\bibinfo {year} {2018})},\ \Eprint
  {http://arxiv.org/abs/1802.00399} {arXiv:1802.00399 [gr-qc]} \BibitemShut
  {NoStop}%
\bibitem [{\citenamefont {Buoninfante}\ \emph {et~al.}(2019)\citenamefont
  {Buoninfante}, \citenamefont {Lambiase},\ and\ \citenamefont
  {Mazumdar}}]{Buoninfante:2018mre}%
  \BibitemOpen
  \bibfield  {author} {\bibinfo {author} {\bibfnamefont {L.}~\bibnamefont
  {Buoninfante}}, \bibinfo {author} {\bibfnamefont {G.}~\bibnamefont
  {Lambiase}}, \ and\ \bibinfo {author} {\bibfnamefont {A.}~\bibnamefont
  {Mazumdar}},\ }\href {\doibase 10.1016/j.nuclphysb.2019.114646} {\bibfield
  {journal} {\bibinfo  {journal} {Nucl. Phys. B}\ }\textbf {\bibinfo {volume}
  {944}},\ \bibinfo {pages} {114646} (\bibinfo {year} {2019})},\ \Eprint
  {http://arxiv.org/abs/1805.03559} {arXiv:1805.03559 [hep-th]} \BibitemShut
  {NoStop}%
\bibitem [{\citenamefont {Modesto}\ and\ \citenamefont
  {Rachwal}(2017)}]{Modesto:2017sdr}%
  \BibitemOpen
  \bibfield  {author} {\bibinfo {author} {\bibfnamefont {L.}~\bibnamefont
  {Modesto}}\ and\ \bibinfo {author} {\bibfnamefont {L.}~\bibnamefont
  {Rachwal}},\ }\href {\doibase 10.1142/S0218271817300208} {\bibfield
  {journal} {\bibinfo  {journal} {Int. J. Mod. Phys.}\ }\textbf {\bibinfo
  {volume} {D26}},\ \bibinfo {pages} {1730020} (\bibinfo {year}
  {2017})}\BibitemShut {NoStop}%
\bibitem [{\citenamefont {Modesto}\ and\ \citenamefont
  {Rachwal}(2014)}]{Modesto:2014lga}%
  \BibitemOpen
  \bibfield  {author} {\bibinfo {author} {\bibfnamefont {L.}~\bibnamefont
  {Modesto}}\ and\ \bibinfo {author} {\bibfnamefont {L.}~\bibnamefont
  {Rachwal}},\ }\href {\doibase 10.1016/j.nuclphysb.2014.10.015} {\bibfield
  {journal} {\bibinfo  {journal} {Nucl. Phys. B}\ }\textbf {\bibinfo {volume}
  {889}},\ \bibinfo {pages} {228} (\bibinfo {year} {2014})},\ \Eprint
  {http://arxiv.org/abs/1407.8036} {arXiv:1407.8036 [hep-th]} \BibitemShut
  {NoStop}%
\bibitem [{\citenamefont {Modesto}\ and\ \citenamefont
  {Rachwal}(2015)}]{Modesto:2015lna}%
  \BibitemOpen
  \bibfield  {author} {\bibinfo {author} {\bibfnamefont {L.}~\bibnamefont
  {Modesto}}\ and\ \bibinfo {author} {\bibfnamefont {L.}~\bibnamefont
  {Rachwal}},\ }\href {\doibase 10.1016/j.nuclphysb.2015.09.006} {\bibfield
  {journal} {\bibinfo  {journal} {Nucl. Phys. B}\ }\textbf {\bibinfo {volume}
  {900}},\ \bibinfo {pages} {147} (\bibinfo {year} {2015})},\ \Eprint
  {http://arxiv.org/abs/1503.00261} {arXiv:1503.00261 [hep-th]} \BibitemShut
  {NoStop}%
\bibitem [{\citenamefont {Dona}\ \emph {et~al.}(2015)\citenamefont {Dona},
  \citenamefont {Giaccari}, \citenamefont {Modesto}, \citenamefont {Rachwal},\
  and\ \citenamefont {Zhu}}]{Dona:2015tra}%
  \BibitemOpen
  \bibfield  {author} {\bibinfo {author} {\bibfnamefont {P.}~\bibnamefont
  {Dona}}, \bibinfo {author} {\bibfnamefont {S.}~\bibnamefont {Giaccari}},
  \bibinfo {author} {\bibfnamefont {L.}~\bibnamefont {Modesto}}, \bibinfo
  {author} {\bibfnamefont {L.}~\bibnamefont {Rachwal}}, \ and\ \bibinfo
  {author} {\bibfnamefont {Y.}~\bibnamefont {Zhu}},\ }\href {\doibase
  10.1007/JHEP08(2015)038} {\bibfield  {journal} {\bibinfo  {journal} {JHEP}\
  }\textbf {\bibinfo {volume} {08}},\ \bibinfo {pages} {038} (\bibinfo {year}
  {2015})},\ \Eprint {http://arxiv.org/abs/1506.04589} {arXiv:1506.04589
  [hep-th]} \BibitemShut {NoStop}%
\bibitem [{\citenamefont {Percacci}(2017)}]{Percacci:2017fkn}%
  \BibitemOpen
  \bibfield  {author} {\bibinfo {author} {\bibfnamefont {R.}~\bibnamefont
  {Percacci}},\ }\href {\doibase 10.1142/10369} {\emph {\bibinfo {title} {{An
  Introduction to Covariant Quantum Gravity and Asymptotic Safety}}}},\
  \bibinfo {series} {100 Years of General Relativity}, Vol.~\bibinfo {volume}
  {3}\ (\bibinfo  {publisher} {World Scientific},\ \bibinfo {year}
  {2017})\BibitemShut {NoStop}%
\bibitem [{\citenamefont {Reuter}\ and\ \citenamefont
  {Saueressig}(2019)}]{Reuter:2019byg}%
  \BibitemOpen
  \bibfield  {author} {\bibinfo {author} {\bibfnamefont {M.}~\bibnamefont
  {Reuter}}\ and\ \bibinfo {author} {\bibfnamefont {F.}~\bibnamefont
  {Saueressig}},\ }\href {https://tinyurl.com/y2bd4ppp} {\emph {\bibinfo
  {title} {{Quantum Gravity and the Functional Renormalization Group}}}}\
  (\bibinfo  {publisher} {Cambridge University Press},\ \bibinfo {year}
  {2019})\BibitemShut {NoStop}%
\bibitem [{\citenamefont {Donoghue}(2020)}]{Donoghue:2019clr}%
  \BibitemOpen
  \bibfield  {author} {\bibinfo {author} {\bibfnamefont {J.~F.}\ \bibnamefont
  {Donoghue}},\ }\href {\doibase 10.3389/fphy.2020.00056} {\bibfield  {journal}
  {\bibinfo  {journal} {Front. in Phys.}\ }\textbf {\bibinfo {volume} {8}},\
  \bibinfo {pages} {56} (\bibinfo {year} {2020})},\ \Eprint
  {http://arxiv.org/abs/1911.02967} {arXiv:1911.02967 [hep-th]} \BibitemShut
  {NoStop}%
\bibitem [{\citenamefont {Bonanno}\ \emph {et~al.}(2020)\citenamefont
  {Bonanno}, \citenamefont {Eichhorn}, \citenamefont {Gies}, \citenamefont
  {Pawlowski}, \citenamefont {Percacci}, \citenamefont {Reuter}, \citenamefont
  {Saueressig},\ and\ \citenamefont {Vacca}}]{Bonanno:2020bil}%
  \BibitemOpen
  \bibfield  {author} {\bibinfo {author} {\bibfnamefont {A.}~\bibnamefont
  {Bonanno}}, \bibinfo {author} {\bibfnamefont {A.}~\bibnamefont {Eichhorn}},
  \bibinfo {author} {\bibfnamefont {H.}~\bibnamefont {Gies}}, \bibinfo {author}
  {\bibfnamefont {J.~M.}\ \bibnamefont {Pawlowski}}, \bibinfo {author}
  {\bibfnamefont {R.}~\bibnamefont {Percacci}}, \bibinfo {author}
  {\bibfnamefont {M.}~\bibnamefont {Reuter}}, \bibinfo {author} {\bibfnamefont
  {F.}~\bibnamefont {Saueressig}}, \ and\ \bibinfo {author} {\bibfnamefont
  {G.~P.}\ \bibnamefont {Vacca}},\ }\href {\doibase 10.3389/fphy.2020.00269}
  {\bibfield  {journal} {\bibinfo  {journal} {Front. in Phys.}\ }\textbf
  {\bibinfo {volume} {8}},\ \bibinfo {pages} {269} (\bibinfo {year} {2020})},\
  \Eprint {http://arxiv.org/abs/2004.06810} {arXiv:2004.06810 [gr-qc]}
  \BibitemShut {NoStop}%
\end{thebibliography}%

\end{document}